\documentclass{emulateapj}
\usepackage{graphicx}

\begin{document}
\title{Probing the Dependence of the Intergalactic Medium on Large Scale Environment Using the Low Redshift Lyman Alpha Forest}
\author{Stephanie Tonnesen$^{1}$, Britton D. Smith$^{2}$, Juna A. Kollmeier$^{1}$, Renyue Cen$^{3}$}
\affil{$^{1}$The Observatories of the Carnegie Institution for Science, 813 Santa Barbara St, Pasadena, CA, 91101 $^{2}$San Diego Supercomputer Center, University of California, San Diego, CA, 92093 $^{3}$Department of Astrophysical Sciences, Princeton University, Peyton Hall, Princeton, NJ, 08544}
\email{1 stonnes@gmail.com (ST)} 

\abstract{We examine the statistics of the low-redshift Ly$\alpha$ forest in an adaptive mesh refinement hydrodynamic cosmological simulation of sufficient volume to include distinct large-scale environments.  We compare our HI column density distribution of absorbers both with recent work and between two highly-refined regions of our simulation:  a large-scale overdensity and a large-scale underdensity (on scales of approximately 20 Mpc).  We recover the average results presented in Kollmeier et al. (2014) using different simulation methods.  We further break down these results as a function of environment to examine the detailed dependence of absorber statistics on large-scale density.  We find that the slope of the HI column density distribution in the 10$^{12.5}$ $\le$ N$_{\rm HI}$/cm$^{-2}$ $\le$ 10$^{14.5}$ range depends on environment such that the slope becomes steeper for higher environmental density, and this difference reflects distinct physical conditions of the intergalactic medium on these scales.  We track this difference to the different temperature structures of filaments in varying environments.  Specifically, filaments in the overdensity are hotter and, correspondingly, are composed of gas with lower HI fractions than those in underdense environments.  Our results highlight that in order to understand the physics driving the HI CDD, we need not only improved accounting of the sources of ionizing UV photons, but also of the physical conditions of the IGM and how this may vary as a function of large-scale environment.}
}
\section{Introduction}
The high-redshift Ly$\alpha$ forest has long been used for cosmology owing to its relatively simple structure theoretically and observationally (e.g., Cen et al. 1994; Zhang et al. 1995; Hernquist et al. 1996; Miralda-Escude et al. 1996; Rauch et al. 1997; Weinberg 1998; Peeples et al. 2010).  This structure, first understood using cosmological hydrodynamics simulations, emerges from the thermodynamic conditions of Hydrogen within the cosmic web and depends only on cosmological parameters and atomic physics through the ionization and recombination rates of Hydrogen.  While the values for cosmological parameters have become exquisitely precise, the same is not true for the metagalactic ionization rate, $\Gamma$, at any redshift.  

Kollmeier et al. (2014; K14) highlighted the use of the low-redshift Ly$\alpha$ forest as a cosmic calorimeter and probe of the low redshift ionizing flux, which is notoriously difficult to measure.  While standard practice at high redshift, K14 found that the metagalactic photoionization rate required by their simulations to match the observed properties of the low-redshift Ly$\alpha$ forest is a factor of five larger than the values predicted by state of the art ionization models of Haardt \& Madau (2012; HM12) --- a discrepancy between models and observations they termed the ``Photon Underproduction Crisis" (PUC).  This is illustrated when comparing the column density distribution (CDD) of Ly$\alpha$ forest absorbers from their simulations, adjusted to different values of $\Gamma$, with data from HST.  Prompted by K14, other authors also found they required higher photoionization rates than HM12 (e.g. Shull et al. 2015; Emerick et al. 2015; Puchwein et al. 2015), although, in Shull et al. (2015) the particular boosts required was a factor of 2-3 rather than 5. In K14, the authors put forward several plausible solutions to the PUC, including increased AGN fractions and ionizing radiation escape fractions relative to the HM12 predictions which they deemed unlikely but have been examined further in the literature (e.g Khaire \& Srianand 2015a,b; who also increase the star formation rate density at $z$ $<$ 0.5 above that of HM12). 

Another possible solution was put forward by Gurvich et al. (2016) using the Illustris simulation including AGN feedback.  They find that for their simulation, the HI CDD in the range N$_{\rm HI}$ $=$ 10$^{12.5}$ to 10$^{14}$ cm$^{-2}$ agrees well with observations due partly to a slightly higher adopted value of $\Gamma$ (Faucher-Gigu{`e}re et al. 2009) but primarily from the inclusion of AGN feedback. The authors note that their simulation begins to diverge from observations at N$_{\rm HI}$ $>$ 10$^{13.5}$ cm$^{-2}$, perhaps indicating excessively strong AGN feedback unphysically reducing the number of absorbers (Gurvich et al. 2016).  

Prompted by this literature, we set out to understand two elements of the PUC which have not been clarified. Firstly what, if any, role does the underlying hydrodynamic scheme play in determining the HI CDD?  One of the possible (but deemed unlikely) resolutions presented for completeness in K14 was the possible effect of the underlying hydrodynamics scheme which could play a role in the PUC.  Those authors quoted results for their smoothed particle hydrodynamics (SPH) simulations which are thought to be robust in the regime of the IGM (certainly at the 50\% level), but it was not directly tested at low redshift whether there could have been an unexpected and far larger effect.  We do that test here.  Secondly, we wished to use the unique nature of our simulation to determine whether the environment would affect the intergalactic medium to an extent that the HI CDD could differ across different regions of the universe.

In this paper we examine Ly$\alpha$ absorbers with HI column densities between 10$^{12.5}$ - 10$^{14.5}$ cm$^{-2}$ in a large-scale ($\sim$20 Mpc) overdensity and underdensity cosmologically simulated using the adaptive mesh refinement (AMR) Eulerian hydrodynamical code \textit{Enzo} (Bryan et al. 2014).  These simulations provide us with an opportunity to compare Ly$\alpha$ absorbers in different large-scale environments within the same fully hydrodynamical cosmological simulation performed at sub-kpc resolution.  Comparing different environments within the same simulation where all physical processes are modeled in the same way minimizes the degrees of freedom of the numerical problem.

In Section \ref{sec:method} we describe our simulations, and briefly mention our galaxy selection technique in Section \ref{sec:galaxyselection}.  We then discuss our methods for creating sightlines through our simulations and spectrally identifying Ly$\alpha$ absorbers in Section \ref{sec:absorptionspectrum}, and explain our method for finding individual Ly$\alpha$ absorbing clouds in Section \ref{sec:absorbers}.  Our first result is a spectrally-determined HI CDD, in Section \ref{sec:HICDD}.  In Section \ref{sec:clouds} we examine individual absorbing clouds we find along our sightlines (Section \ref{sec:cloudproperties}), the gas properties of those clouds (Section \ref{sec:cloudgas}) and where the clouds tend to be found (Section \ref{sec:cloudpositions}).  In Section \ref{sec:cloudenvironment} we synthesize our absorbing cloud results to determine the impact of environment on Ly$\alpha$ clouds.  Finally, in Section \ref{sec:conclusions} we summarize our conclusions, and discuss directions for both observational and theoretical future work.

\section{Method}\label{sec:method}

For the details of our simulations, we refer the reader to Cen (2012a), although for completeness we reiterate the main points here.  We perform cosmological simulations with the AMR Eulerian hydrodynamical code \textit{Enzo} (Bryan 1999; Bryan et al. 2014).  We use cosmological parameters consistent with the WMAP7-normalized LCDM model (Komatsu et al. 2011):  $\Omega_M$ = 0.28, $\Omega_b$ = 0.046, $\Omega_\Lambda$ = 0.72, $\sigma_8$ = 0.82, $H_o$ = 100 $h$ km s$^{-1}$ Mpc$^{-1}$ = 70 km s$^{-1}$ Mpc$^{-1}$, and $n$ = 0.96.  We first ran a low resolution simulation with a periodic box of 120 $h^{-1}$ Mpc on a side, and identified two regions:  an overdensity centered on a cluster and an underdensity centered on a void at $z = 0$.  We then resimulated each of the two regions separately with high resolution, but embedded within the outer 120 $h^{-1}$ Mpc box to properly take into account large-scale tidal field effects and appropriate fluxes of matter, energy and momentum across the boundaries of the refined region.

The overdense refined region is 21 $\times$ 24 $\times$ 20 $h^{-3}$ Mpc$^3$.  The central cluster is $\sim$2 $\times$ 10$^{14}$ M$_\odot$ with a virial radius (r$_{200}$) of 1.3 $h^{-1}$ Mpc.  The underdense refined region is somewhat larger, at 31 $\times$ 31 $\times$ 35 $h^{-3}$ Mpc$^3$.  At their respective volumes, they represent +1.8$\sigma$ and -1.0$\sigma$ fluctuations.  Although these are large-scale over- and underdense environments, there are galaxies at a range of local densities in both boxes, and there is substantial overlap of local densities between the two volumes (Tonnesen \& Cen 2012).

In both refined boxes, the minimum cell size is 0.46 $h^{-1}$ kpc, using 11 refinement levels at $z = 0$.  The initial conditions for the refined regions have a mean interparticle separation of 117 $h^{-1}$ kpc comoving, and a dark matter particle mass of 1.07 $\times$ 10$^8$ $h^{-1}$ M$_\odot$.  While we do not perform a resolution study here, we refer our readers to Tepper-Garc{\'{\i}}a et al. (2012).  They show that the HI column density distribution between 10$^{12.5}$ $\le$ N$_{\rm HI}$/cm$^{-2}$ $\le$ 10$^{14.5}$, one of the major results of our paper, is well converged using dark matter particle masses of 4.1 $\times$ 10$^8$ $h^{-1}$ M$_\odot$ (a factor of 4 more massive than the dark matter particle mass in our simulation).

The simulations include a metagalactic UV background (Haardt \& Madau 1996) supplemented with an X-ray Compton heating background from Madau \& Efstathiou (1999), a model for shielding of UV radiation by neutral hydrogen (Cen et al. 2005), and metallicity-dependent radiative cooling (Cen et al. 1995).  The fraction and density of neutral hydrogen is directly computed within the simulations.  Star particles are created in gas cells that satisfy a set of criteria for star formation proposed by Cen \& Ostriker (1992), and reiterated with regards to this simulation in Cen (2012a).  Each star particle has a mass of $\sim$10$^6$ M$_\odot$, which is similar to the mass of a coeval globular cluster.  Once formed, the stellar particle loses mass through gas recycling from Type II supernovae feedback, and about 30\% of the stellar particle mass is returned to the ISM within a time step.  Supernovae feedback is implemented as described in Cen (2012a):  feedback energy and ejected metal-enriched mass are distributed into 27 local gas cells centered at the star particle in question, weighted by the specific volume
of each cell.  We allow the whole feedback process to be hydrodynamically coupled to surroundings and subject to
relevant physical processes, such as cooling and heating, as in
nature.  The simulation used in this paper has compared several galaxy properties that depend critically on the feedback method to observations and found strong agreement (Cen 2012a-b; Cen 2013).  For example, Cen (2012a) found excellent agreement between simulated and observed damped Ly$\alpha$ systems, and Cen (2012b) found that the properties of O VI and O VII absorbers also agree well with observations.  We do not include a prescription for AGN feedback in this simulation, and as a result, our simulation overproduces luminous galaxies at the centers of groups and clusters of galaxies.  

\subsection{Galaxies}\label{sec:galaxyselection}

As has been discussed in earlier work (e.g. Tonnesen \& Cen 2012; 2014; 2015), we use HOP (Eisenstein \& Hut 1998) to identify galaxies using the stellar particles.   HOP uses a two-step procedure to identify individual galaxies. First, the algorithm assigns a density to each star particle based on the distribution of the surrounding particles and then hops from a particle to its densest nearby neighbor until a maximum is reached. All particles (with densities above a minimum threshold, $\delta_{outer}$) that reach the same maximum are identified as one coherent group.  The densest star particle is considered the center of the stellar group, or galaxy.  In the second step, groups are combined if the density at the saddle point which connects them is greater than $\delta_{saddle}$.  We use HOP because of its physical basis, although we expect similar results would be found using a friends-of-friends halo finder.  HOP has been tested and shown to be robust (e.g. Tonnesen, Bryan, \& van Gorkom 2007).  

\subsection{Sightlines and Spectra}\label{sec:absorptionspectrum}

We follow the general method described in K14 to calculate sightlines and spectra through our simulated regions.  We use the $z$=0.1 outputs of both refined region simulations.  In each refined box, we then extract 2500 sightlines that span 0.008 in redshift.  This width spans a large fraction of the refined region while ensuring that the same structures are not sampled repeatedly in a sightline.  

Specifically, we use yt\footnote{http://yt-project.org/} to create the sightlines and spectra across our refined regions.  yt is a python package for analyzing and visualizing volumetric, multi-resolution data from astrophysical simulations and observations (Turk et al. 2011).  We randomly generate sightlines that are completely contained within the refined region of our simulation, allowing the rays to wrap around the refined region.  In the overdense region, we only allow our sightlines to span 21 $\times$ 18 $\times$ 20 $h^{-3}$ Mpc$^3$ so that they are crossing truly highly-refined structures.  We generate a synthetic spectrum for each ray and derive HI column densities from the spectrum using the methods described below.

For creating an absorption spectrum from the simulation data, we use the absorption spectrum generator in yt (Turk et al. 2011). We create artificial sight lines from simulation data by extracting all grid cells intersected the ray.  We gather all relevant field data for each grid cell, including density fields for appropriate ions, velocities, and the temperature.  Spectra are generated by depositing a Voigt profile for each element of the ray, where the column density is the density of relevant ion multiplied by the path length through each cell and the Doppler parameter is given by the cell temperature.  The lines are shifted from their rest wavelengths by a combination of the line-of-sight peculiar velocity and Hubble expansion.  For the redshift, the start of the ray is assumed to be at the redshift of the dataset and the redshift is increased according to the comoving radial distance along the ray.  In practice, spectral features identified as single lines are the result of multiple elements in the ray.  This functionality has now been subsumed by the Trident package (Hummels et al. 2016), and we refer the reader there for a detailed description of the method.  Because we are only interested in Ly$\alpha$ in this work, we only include the Ly$\alpha$ line in the synthetic spectra and assume zero noise.  While unrealistic, this sets an upper limit on the number of detectable lines.  As the simulation includes non-equilibrium atomic H/He chemistry, we are able to provide the neutral H densities directly to the spectrum generator.  Automated spectral fitting is done using the method of Egan et al. (2014, E14), also a part of the yt package.  This method uses the least squares fitting algorithm provided by the SciPy “optimize” library, to find an optimal combination of column density, Doppler parameter, and redshift for any contiguous region of a spectrum below 99\% of the continuum flux.  Additional components are added, up to a maximum of 8, if the reduced $\chi^{2}$ error of the optimal fit is above 10$^{-4}$.  To aid the fitting routine, we allow for only a range of acceptable parameters, (10$^{11}$ $<$ N$_{\rm HI}$ $<$ 10${22}$ cm$^{-2}$), Doppler parameter (1 $<$ b $<$ 300), and redshift (0.1 $<$ $z$ $<$ 0.4).  These settings are broader than the range of parameters of the fitted lines.

This method can be compared directly to observations, because it uses the redshift of Ly$\alpha$ absorbing gas to determine whether it is combined into a single absorption feature or separated into several features.  This procedure thus closely resembles the procedure followed by observers when analyzing real data.  

\subsection{Selecting Individual Clouds}\label{sec:absorbers}

In order to examine the properties of individual absorbers along a sightline, we select dense HI clouds, using a method similar to the `contour method' described in E14.  To do this, we find local maxima in HI number density by selecting a cell with the peak HI density within $\pm$ 300 kpc from that point.  This peak HI number density must be greater than 10$^{-13}$ cm$^{-3}$ (although as we discuss below, this requirement does not affect our results).  We then select the edges of the absorbers by finding the closest cell that is either a local minimum within $\pm$ 300 kpc or has an HI number density that is n$_{\rm HI,min}$/n$_{\rm HI,peak}$ = \{0.1,0.5,0.9\} of the peak value.  We use the maximum or minimum found over several cells, here selected to be $\pm$ 300 kpc, to smooth out small density fluctuations between cells, even in regions far from galaxies that are not refined to the maximum level.  As a straightforward check of our method, we plotted the HI density profiles of several absorbers and found that from peak to minimum they were $\geq$ 250 kpc, which means that the minimum absorber size was not likely dictated by the criterium that there be $>$300 kpc between neighboring minima (or maxima).  Multiple peaks are rarely seen in our visual inspection of HI absorber profiles, so we are identifying single clouds.  

\textit{We will call absorbers selected in this manner clouds throughout the paper in order to differentiate them with absorbers found using the spectrum as described above.} 

%fig1
\begin{figure}
\includegraphics[scale=0.64]{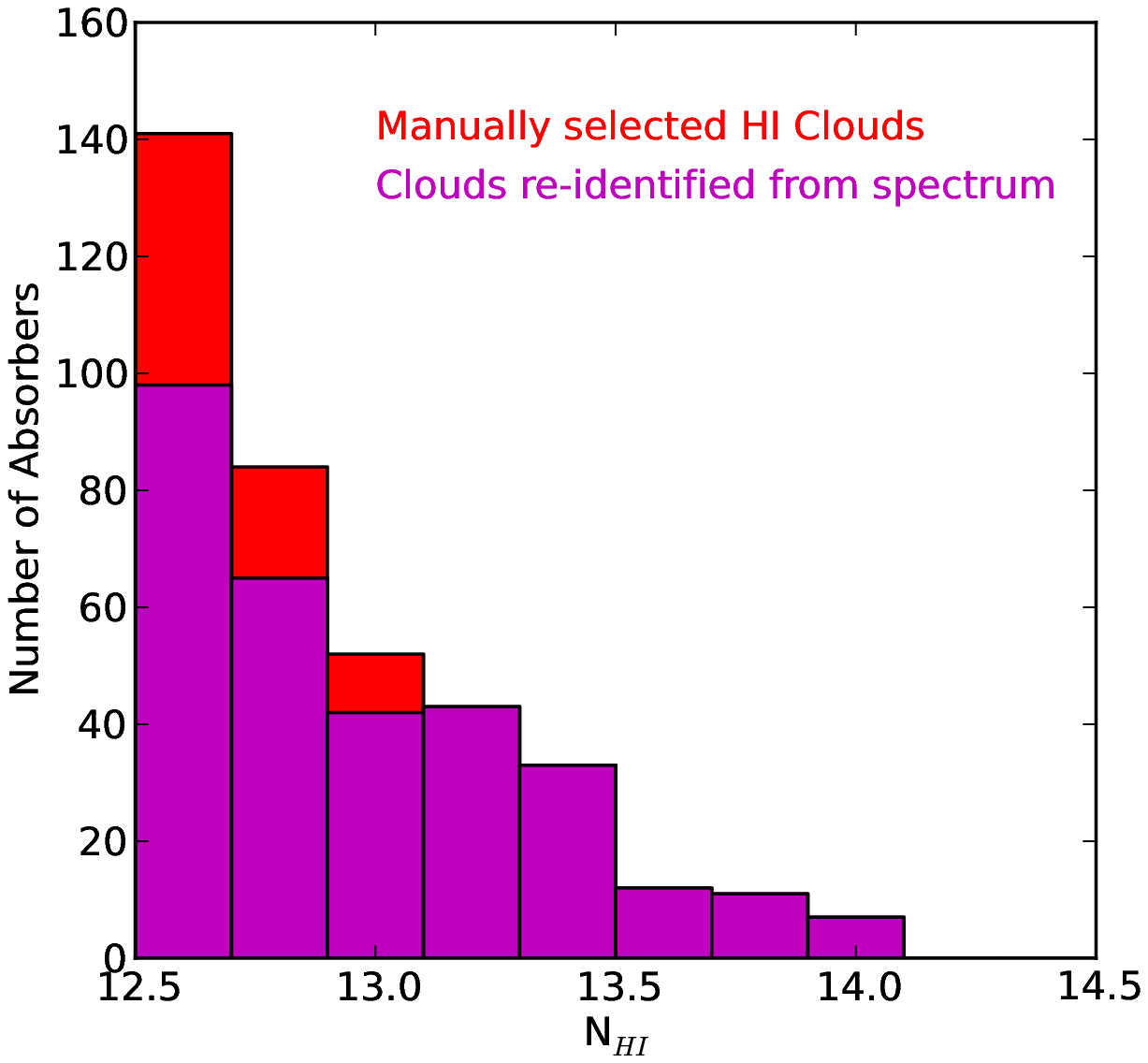}\\%{absorbercheck_CBOX_NH_histogram.eps}
\includegraphics[scale=0.64]{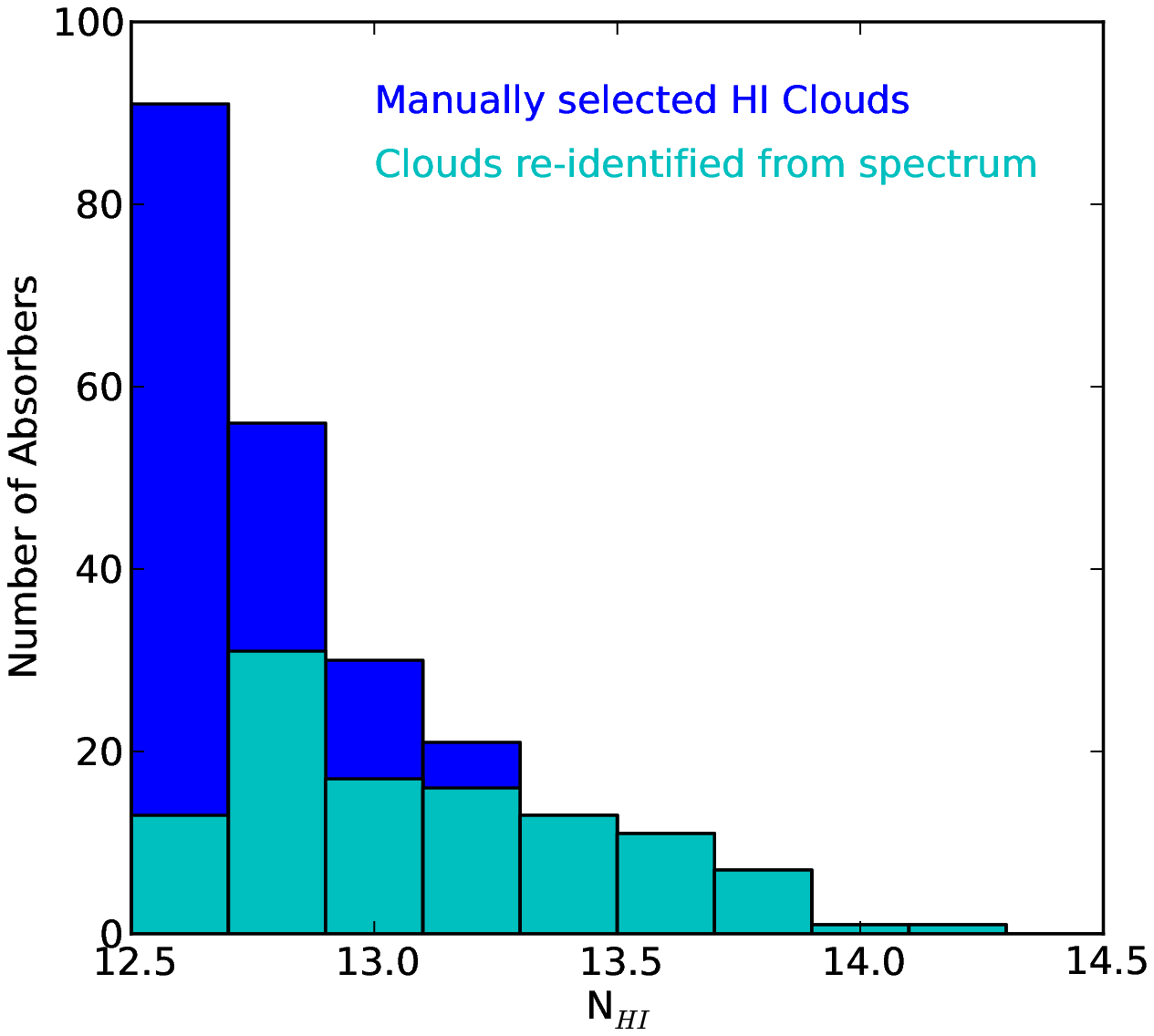}%{absorbercheck_VBOX_NH_histogram.eps}
\caption{Histograms comparing manually selected HI clouds and those that are subsequently identified as Ly$\alpha$ absorbers by creating an absorption spectrum using yt.  200 sightlines were used in this comparison in both the overdense (top panel) and underdense (bottom panel) refined regions.  \textbf{Clouds selected in either manner show the same difference in the HI CDD between different environments.}} \label{fig:absorbercheck}
\end{figure}

Because we are not including redshift information in our cloud selection and allow for broad absorbers with little variation between the peak and minimum HI density, we do not expect that every cloud will be identified using the yt absorption spectrum routine.  Indeed, this is what we find when compare our two routines.  To do this, we first select individual HI clouds along 200 lines of sight.  Then for each cloud, we run the yt absorption spectrum generator routine on the small section of the sightline that contains the cloud.  Thus we can determine whether the yt routine would identify each individual HI cloud.  We show in Figure \ref{fig:absorbercheck} the histograms of clouds that are identified using n$_{\rm HI,min}$/n$_{\rm HI,peak}$ = 0.1, and overplot the number of those clouds that are re-identified using yt to create a spectrum of that section of the sightline and identifying the Ly$\alpha$ line as described in Section \ref{sec:absorptionspectrum}.  We find reasonable agreement in the HI cloud column densities measured using both methods.  In both the overdense and underdense regions, $\sim$85\% of clouds that are re-identified with the yt routine have absorption spectrum column densities within 20\% of the HI cloud column density.  In the overdense (underdense) region about 7\% (16\%) of the re-identified HI clouds have multiple spectral peaks.  There is no clear reason for the differences between cloud re-identification in the two environments.  We find that the number of HI clouds re-identified with the yt routine does not change when we vary the fitting parameters discussed in Section \ref{sec:absorptionspectrum} (for example, the initial guess for the fitting routine of the column density or b parameter, or changing the range of allowed column densities or redshifts).  However, if we only select HI clouds whose peak HI number density is greater than 9$\times$10$^{-12}$ cm$^{-3}$ (which corresponds to a neutral fraction of 1.4 $\times$ 10$^{-5}$ at three times the mean baryon density), then in the overdense and underdense boxes, respectively, 99\% and 82\% of clouds are identified as Ly$\alpha$ absorbers using the yt absorption spectrum generator routine.  We vary the peak number density simply to illustrate that, as we expect, HI clouds with lower peak densities are more difficult for the yt absorption spectrum routine to identify.

Whether we use our fiducial peak HI number density minimum, this high minimum, or no minimum, the comparisons we discuss in this paper remain qualitatively the same.  

We make use of both our spectrally identified absorbers as well as the manually identified clouds.  The spectrally identified absorbers are most useful for comparing with observational datasets.  However, for exploring the physical conditions giving rise to the absorption, and thus to gain further physical insight into the environmental dependence of absorption, we will use the statistics from our manually identified cloud population.  Comparing Figures \ref{fig:HI_CDD} and \ref{fig:Nabsorbers}, we find that either absorber selection method finds that the distribution of HI column densities differs between the overdense and underdense environments.  We first turn our attention to comparison with observations and prior work.

%\section{Results}
\section{The HI Column Density Distribution}\label{sec:HICDD}

%fig2
\begin{figure}
\includegraphics[scale=0.63]{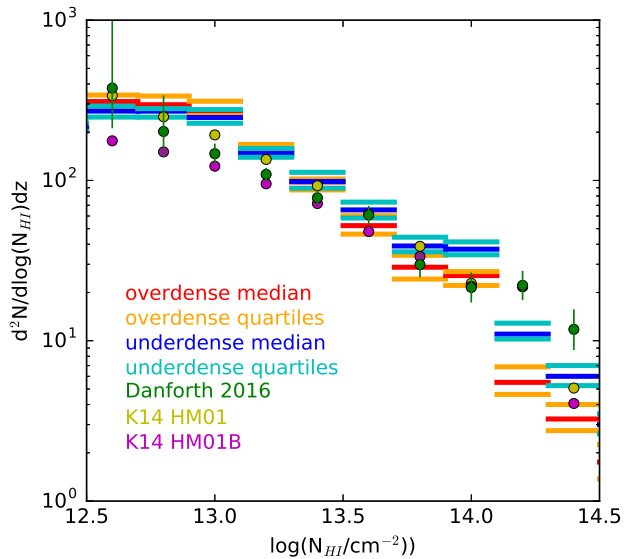}%{HI_CDD_2500_compare_percentilesonly_smallyCBOX_smalldz.eps}
\caption{The HI CDD for our sightlines.  We generated 15 sets of 2500 sightlines with $\Delta$z= 0.008. The red and blue lines are the median number of absorbers in the overdense and underdense regions, respectively.  The orange and cyan lines denote the upper and lower quartiles for the overdense and underdense regions.  The HI CDD in the overdense region is steeper than in the underdense region.\label{fig:HI_CDD}}
\end{figure}

We first create an HI Column Density Distribution (CDD) for comparison with K14 and observations.  As in K14, this is defined to be the mean number of absorbers per logarithmic interval of column density per unit redshift path length, for the simulated Ly$\alpha$ forest.  In the z=0.1 output of both of our refined regions, we generated 15 sets of 2500 sightlines with $\Delta$z= 0.008.  We then identified absorbers in each of those 15 sets of sightlines using the technique described in Section \ref{sec:absorptionspectrum}.

In Figure \ref{fig:HI_CDD} we show the resulting HI CDD for each refined region.  In blue and cyan we show the median and quartile range, respectively, of our 15 sets of sightlines in the underdense region, and in red and orange we show the median and quartile range, respectively, of our 15 sets of sightlines in the overdense region.  We have also included two of the K14 HI CDDs, specifically those that use the Haardt \& Madau (2001) UV background (HM01; yellow) and the Haardt \& Madau (2001) UV background plus blazars (HM01B; magenta).  In green we include the observational results from Danforth et al. (2016).  

First, we find that in general we show broad agreement with observational results, particularly at column densities below 10$^{14}$ cm$^{-2}$, as well as with the K14 HM01 prediction as discussed in K14.  We note, however, that our UV background is a hot HM01, or HM01 plus an X-ray Compton heating background, which produces a systematically ``thinner" forest than the more recent Haardt \& Madau (2012) and Faucher-Gigu{'e}re et al. (2009) UV backgrounds (see Figure 1 in K14).  The HM01 UV background is significantly larger than some more recent direct measurements of the local metagalactic UVB (Adams et al. 2011).  We also note that the slope of our CDD in either refined region is steeper than observed by Danforth et al. (2016), particularly at high columns.  

Our simulation is unique in that we can compare two highly-resolved regions around a large-scale overdensity and underdensity.  We note that the effect of the different environment becomes manifest at columns greater than $10^{13.5}$ cm$^{-2}$ and that below this column, the effect of large-scale overdensity is less obvious, although still distinct. We find that the HI CDD in the overdense region is {\it steeper} than in the underdense region.  This is driven partly by the absence of low-column density (N$_{\rm HI} < 10^{13.5}$ cm$^{-2}$) absorbers in underdense void regions, but more clearly by the absence of high-column density N$_{HI} > 10^{13.5}$ absorbers in the overdense volume.

We note that we find the same qualitative results when we vary the simulation data we collect.  For example, our results did not differ if we used 50 sets of 2500 sightlines or 7 sets of 2500 sightlines, if we increased the $\Delta$z of each sightline to 0.016, or if we used a larger overdense box that included some underrefined regions.  

\section{Examining Absorbing Clouds in Detail}\label{sec:clouds}

\begin{figure}
\includegraphics[scale=0.63]{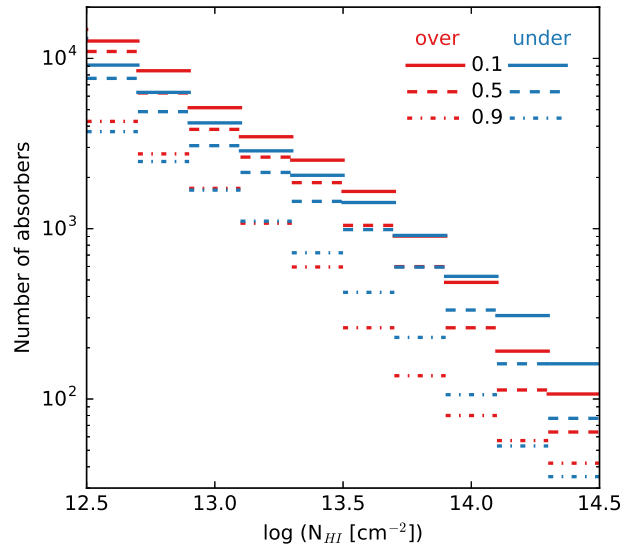}%{Nabsorbers_vlos_ranges_toh_nr.eps}
\caption{The number of HI clouds selected by finding a local n$_{\rm HI}$ maximum and calculating the total column density between cells with \{0.1,0.5,0.9\} of the peak value.  The overdense box is denoted by red lines, and the underdense box is denoted by blue lines.  The linestyle, \{solid, dashed, dash-dot\} designates the HI minimum of the clouds: \{0.1,0.5,0.9\}$\times$n$_{peak}$.  HI clouds selected this way show a steeper slope in the number of absorbers as a function of N$_{\rm HI}$ in the overdense box than in the underdense box for the clouds that include gas with densities down to  \{0.1,0.5\} of the peak value.  \label{fig:Nabsorbers}}
\end{figure}

We now examine HI fabsorbers in these two regions to determine whether there are differences in their properties that could drive the differences in the HI CDD.  In order to determine if cloud properties differ in the two environments, we identify individual HI clouds as we discuss in Section \ref{sec:absorbers}.  For the following results we use all of the HI clouds identified in 7 sets of 2500 lines of sight.  As we have shown above, our HI cloud selection scheme does not correspond directly to spectral absorbers.  We first create a CDD of all of the HI clouds found in these 17,500 lines of sight.  Figure \ref{fig:Nabsorbers} shows this CDD for the three cloud definitions in each of the refined regions.  The overdense box is denoted by red lines and the underdense box is denoted by blue lines.  The linestyle designates whether the clouds are \{0.1,0.5,0.9\}$\times$n$_{peak}$. 

As with the spectrally-determined HI CDD in Figure \ref{fig:HI_CDD}, the slope in the overdense region is steeper than the slope in the underdense region for all but our smallest clouds (n$_{\rm HI,min}$/n$_{\rm HI,peak}$ = 0.9).  Therefore, looking directly at the HI clouds should provide physical insight into environmental causes that may drive the differences in the spectrally-observed HI CDDs.

%fig3

\subsection{Cloud Size}\label{sec:cloudproperties}

\begin{figure}
\includegraphics[scale=0.55]{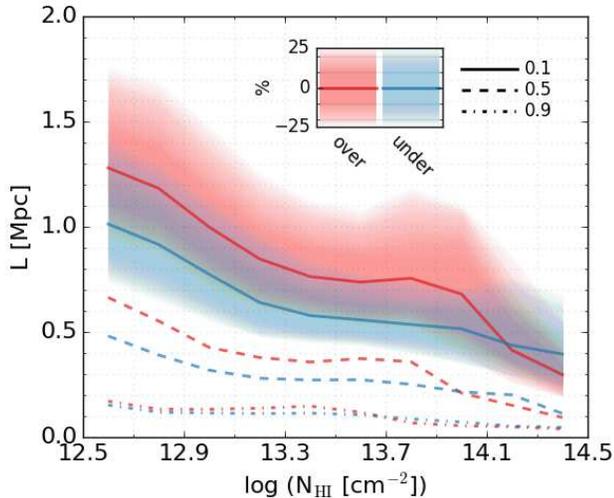}%{L_avg.eps}%{Lmpc_vlos_ranges_toh_nr.png}
\caption{The length of HI clouds, binned by N$_{\rm HI}$.  The colors are the same as in Figure \ref{fig:Nabsorbers}.  The lines denote the median value in a bin and the shaded region spans the quartile range of the n$_{\rm HI,min}$/n$_{\rm HI,peak}$ = 0.1 clouds.    \label{fig:Lmpc}}
\end{figure}

In Figure \ref{fig:Lmpc} we plot the length of HI clouds, binned by N$_{\rm HI}$.  The colors are as in Figure \ref{fig:Nabsorbers}, with the lines denoting the median values and the shaded regions spanning the upper to lower quartiles of the n$_{\rm HI,min}$/n$_{\rm HI,peak}$ = 0.1 clouds.  The larger clouds (\{0.1,0.5\}$\times$n$_{peak}$) have consistently broader lengths in the overdense region up to the highest column densities (log(N$_{\rm HI}$/cm$^{-2}$) $\ge$ 14).  However, the differences between the cloud sizes are not dramatic, and there is a large overlap in the range of cloud size values, as shown by the shaded quartile regions.  The smallest clouds, n$_{\rm HI,min}$/n$_{\rm HI,peak}$ = 0.9, are nearly identical in size in the two environments across the entire column density range.

We can compare our cloud lengths to E14.  We find that our largest absorbers, those with n$_{\rm HI,min}$/n$_{\rm HI,peak}$ = 0.1, tend to be larger than the E14 clouds at all column densities.  Clouds selected with n$_{\rm HI,min}$/n$_{\rm HI,peak}$ = \{0.5, 0.9\} are larger at lower column densities and smaller at higher column densities than E14 clouds.  Part of this is that E14 uses a density cutoff of n$_{\rm HI,peak}$ $\ge$ 10$^{-12}$.  Indeed, we find, unsurprisingly, that increasing our density cutoff for the peak value also decreases our absorber sizes, as you need a smaller pathlength of higher density gas to reach the same total column density.  However, we reiterate from Section \ref{sec:absorbers} that our relative relationships between absorbers in the two environments do not change depending on our n$_{\rm HI,peak}$ cutoff and it is the relative environmental dependence we wish to probe.  

%fig4

\subsection{The Gas Properties of Absorbing Clouds}\label{sec:cloudgas}

%fig5
\begin{figure}
\includegraphics[scale=0.55]{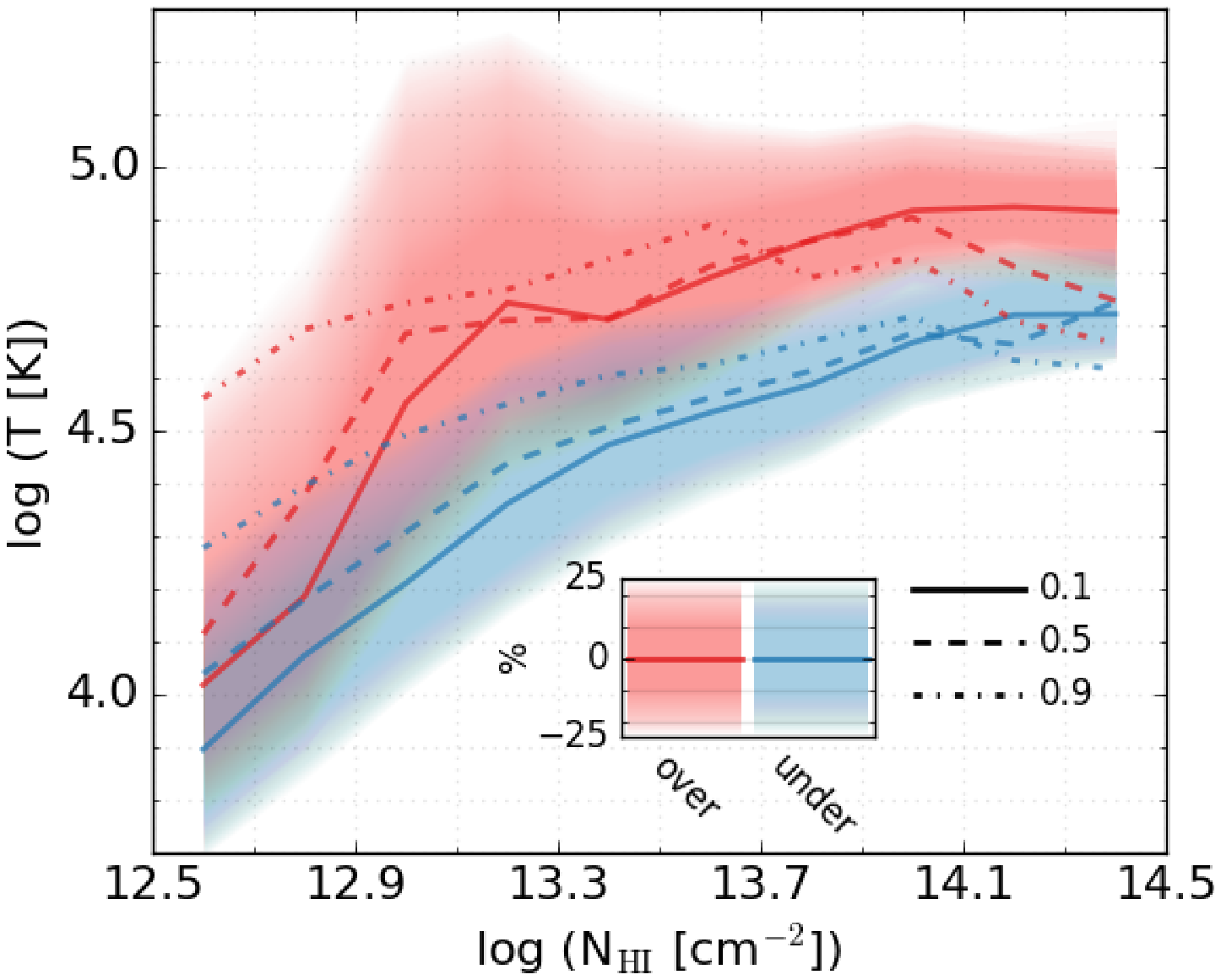}\\%{t_avg.eps}\\%{temp_avg_ranges_toh_all_nr.png}\\
\includegraphics[scale=0.55]{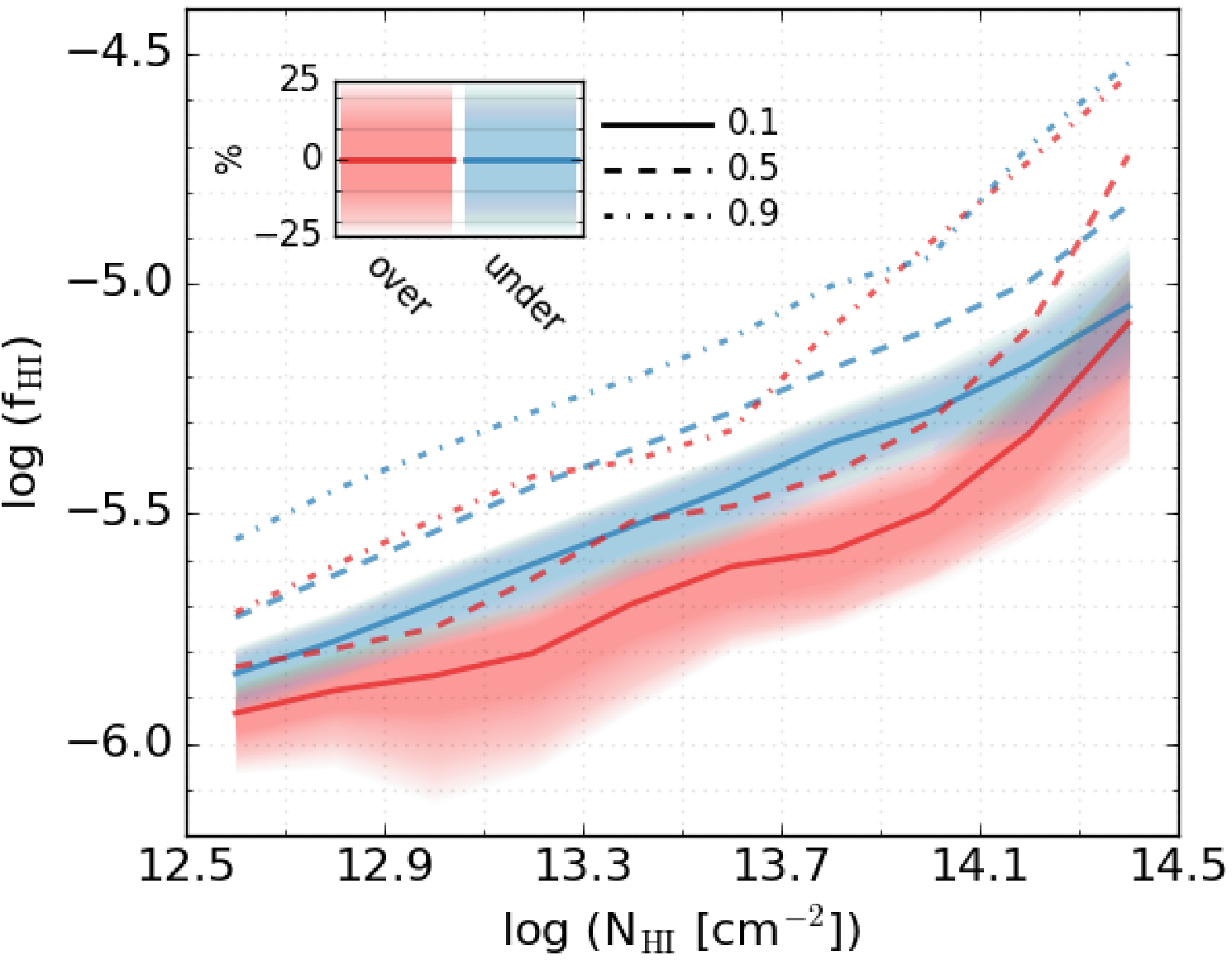}%{fh_avg.eps}%{fh_avg_numd_ranges_toh_all_nr.png}
\caption{Gas properties of absorbing clouds, binned by N$_{\rm HI}$.  The colors are the same as in Figure \ref{fig:Nabsorbers}.  The lines denote the median value in a bin and the shaded region span the quartile range of the n$_{\rm HI,min}$/n$_{\rm HI,peak}$ = 0.1 clouds. \textbf{Top panel:}  The average temperature of absorbing clouds, binned by N$_{\rm HI}$.   The temperature of clouds in the overdense region is consistently higher than those of clouds in the underdense region.  \textbf{Bottom panel:} The average HI fraction of absorbing clouds, binned by N$_{\rm HI}$.   The HI fraction of clouds in the overdense region is consistently lower than those of clouds in the underdense region.  \label{fig:temp}}
\end{figure}

Differences in the sizes of clouds that have the same HI column density may reflect intrinsic differences in the gas that forms these Ly$\alpha$ absorbing clouds.  We now examine this directly.  In the top panel of Figure \ref{fig:temp} we show the average cloud temperature binned by N$_{\rm HI}$.  Our results are the same if we use the median temperature or an average temperature weighted by cell density or column density.  When comparing clouds selected using the same density criteria (the same n$_{\rm HI,min}$/n$_{\rm HI,peak}$ value), the temperature of the clouds in the overdense region is always higher than the temperature of the clouds in the underdense region.  In most bins the clouds in the overdense region are hotter by a factor of at least 1.5.  In fact, for clouds with log(N$_{\rm HI}$/cm$^{-2}$) $>$ 12.9 the density criteria does not matter, and clouds in the overdense region are hotter than clouds in the underdense region.  We might expect hotter clouds to be larger, and indeed, as we have shown, clouds in the overdense region tend to be larger than clouds in the underdense region.  We find that the sizes of our n$_{\rm HI,min}$/n$_{\rm HI,peak}$ $=$ 0.1 clouds are broadly consistent with straightforward Jeans-length arguments that scale with the thermal velocity (e.g. eqn. 2 in Peeples et al. 2010) and indicate a typical size of $\sim$ 800 $h^{-1}$kpc for IGM temperatures of $10^4 K$, with cloud lengths generally ranging from 0.6-1.4 Jeans lengths.

In the bottom panel of Figure \ref{fig:temp} we plot the average HI fraction in the absorbing clouds binned by N$_{\rm HI}$.  As with the temperature measurement, when comparing clouds selected using the same density criteria, the HI fraction of the clouds in the overdense region is always lower than the HI fraction of the clouds in the underdense region.  However, we note that the HI fractions of the smallest clouds (n$_{\rm HI,min}$/n$_{\rm HI,peak}$ $=$ 0.9) are quite similar at higher column densities.  Because HI clouds tend to be hotter in the overdense region, it follows that they would have lower HI fractions.  This also explains the size difference in clouds in the different environments:  hotter clouds have lower neutral fractions, so must be larger to reach the same column density.  

\begin{figure}
\includegraphics[scale=0.55]{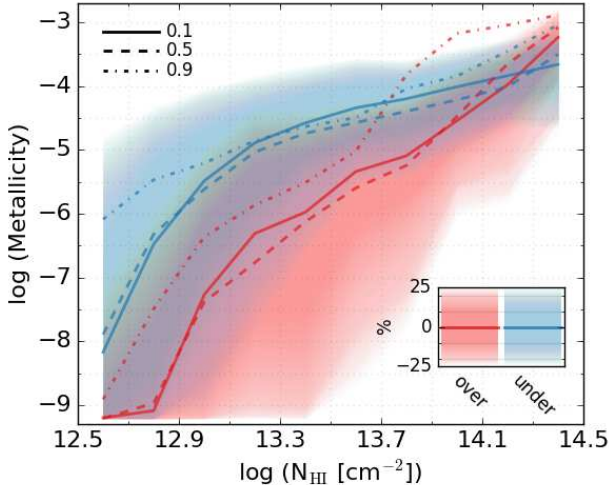}%{met_avg.eps}%{met_avg_ranges_toh_nr.png}
\caption{The average metallicity of HI clouds, binned by N$_{\\rm HI}$.  The colors are the same as in Figure \ref{fig:Nabsorbers}.  The lines denote the median value in a bin and the shaded region spans the quartile range.  The metallicity of clouds in either environment is quite low.  \label{fig:metallicity}}
\end{figure}

These HI absorbing clouds could be formed from gas that has been ejected from galaxies due to feedback or from gas that has never been impacted by outflows.  In order to determine which of these scenarios is more likely, we examine the gas metallicity of the absorbing clouds.  In Figure \ref{fig:metallicity}, we plot the average metallicity of our absorbing clouds binned by log(N$_{\rm HI}$/cm$^{-2}$).  In both the overdense and underdense region, the metallicities are quite low until at least log(N$_{\rm HI}$/cm$^{-2}$) $>$ 13.9.  Below this column density, even the upper quartile of all clouds does not reach a metallicity of 10$^{-3}$.  Even at higher column densities the median value of metallicity remains below 10$^{-3}$ in all clouds except those in the highest bin selected using (n$_{\rm HI,min}$/n$_{\rm HI,peak}$ = 0.9) in the overdense environment.  

Our clouds are largely very low metallicity, so would generally not result in metal detections in observations.  This is roughly in agreement with available observations.  Stocke et al. (2006) find that all of their O VI detections are within 1.15 Mpc of the nearest galaxy in regions surveyed to at least L$^*$, but even within this close distance only $\sim$30\% of Ly$\alpha$ absorbers have O VI detections.  Many of our clouds are more distant that 1.15 Mpc from any galaxy in our simulation (which identifies galaxies down to below 10$^9$ M$_{\odot}$).  In fact, most of our low-column density clouds are beyond 1.15 Mpc from the nearest galaxy (log(N$_{\rm HI}$/cm$^{-2}$) $<$ 13.1).  Danforth et al. (2016) do not give distance information for their observed Ly$\alpha$ absorbers, but find that 3\% of 12.5 $<$ log(N$_{\rm HI}$/cm$^{-2}$) $<$ 13.5 Ly$\alpha$ absorbers have metal detections, and this fraction rises to only 22\% for 13.5 $<$ log(N$_{\rm HI}$/cm$^{-2}$) $<$ 14.5 Ly$\alpha$ absorbers.  As discussed by several authors (eg. Werk et al. 2013, 2014; Danforth et al. 2016), the drop in detection rate with decreasing column density may be because absorber metallicity decreases with decreasing column density, as we find, or because the same metal fraction is increasingly difficult to detect at low column densities.

Low metallicity indicates that in most cases, the gas creating the absorbers has not recently been ejected from galaxies.  Therefore this gas has either had very little interaction with outflows, or was ejected at early times and has thoroughly mixed with the surrounding intergalactic medium (as discussed in Tepper-Garc{\'{\i}}a et al. 2012).  This is particularly interesting when we compare absorbing clouds in the two environments.  Although all of the clouds have low metallicity, in all but the highest column density bins the median metallicity of the absorbing clouds in the overdense box is lower than the median metallicity of clouds in the underdense box.  When we combine this information with the fact that the average temperature of clouds in the overdense box is higher than the temperature of those in the underdense box (Figure 
\ref{fig:temp}), we conclude that these clouds are likely to have been shock heated through structure formation rather than galactic feedback.

%fig6

\subsection{Where are Absorbing Clouds?}\label{sec:cloudpositions}

In order to create a complete picture of absorbing clouds, we must determine where they reside with respect to galaxies.  We calculate the distance to the nearest HOP-identified galaxy (see Section \ref{sec:galaxyselection}).  We also calculate the distance to the nearest central galaxy, which is defined as a galaxy that is not within two r$_{200}$ of any more massive galaxy, using the stellar mass as identified with HOP.  In order to find the smallest distance to any possible galaxy, we do not require a minimum stellar mass for our galaxies.  Our results are qualitatively the same whether or not we include satellite galaxies, so we will focus on the distance to central galaxies.    

%fig7
\begin{figure}
\includegraphics[scale=0.55]{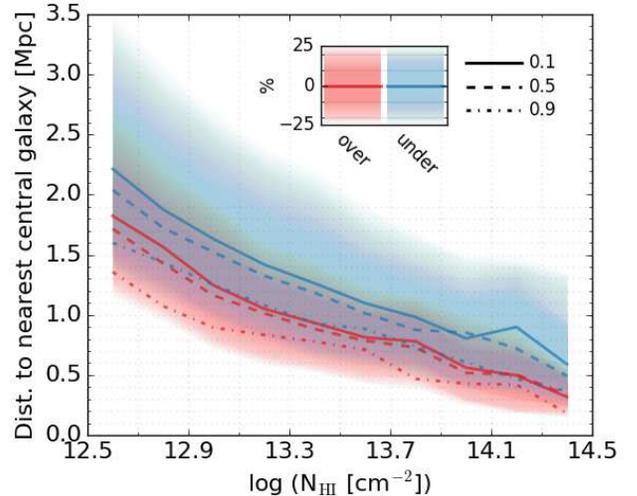}%{galdist.eps}%{dmin_centralgals_sat2_vlos_ranges_toh_nr.png}
\caption{The distance between absorbing clouds and the nearest central galaxy, binned by N$_{\rm HI}$.  The colors are the same as in Figure \ref{fig:Nabsorbers}, although for clarity we only show the largest clouds (n$_{\rm HI,min}$/n$_{\rm HI,peak}$ = 0.1).  The lines denote the median value in a bin and the shaded region spans the quartile range.  All of our clouds are quite distant from the nearest galaxy. \label{fig:gad}}
\end{figure}

In Figure \ref{fig:gad} we plot the distance between absorbing clouds and the nearest central galaxy, binned by N$_{\rm HI}$.  We see that in any N$_{\rm HI}$ bin, the absorbing clouds in the overdense region are closer to galaxies than clouds in the underdense region.  This is expected as there are many more galaxies in the overdense region.  It is notable that the median distance for clouds in all but the highest column density bin is more than 500 kpc.  This is in qualitative agreement with the observational finding of the distances between Ly$\alpha$ absorbers and nearby galaxies by Stocke et al. (2006).  We find that the median stellar masses of the nearest galaxies in any N$_{\rm HI}$ bin are below 2$\times$10$^{10}$ M$_{\odot}$ in either environment, and thus the absorbers tend to be more than 2 virial radii from the nearest galaxy.  In fact, the median mass of galaxies closest to the absorbing clouds is low enough that they may not even produce a strong accretion shock (Dekel \& Birnboim 2006; Keres et al. 2005).  The Ly$\alpha$ forest is formed well outside the circumgalactic medium of galaxies.  

In the overdense region, we also specifically checked the distance between absorbing clouds and the cD galaxy at the center of the most massive cluster in the simulation.  We find that 75\% of the clouds are more than 5 Mpc from the cD galaxy, or more than 2.5 r$_{vir}$ from the cluster center.  The environment within the cluster is not determining the properties of absorbing clouds within the overdense region.

%fig8
\begin{figure*}[!htb]
\includegraphics[scale=0.37,trim=10mm 5mm 49mm 0mm, clip]{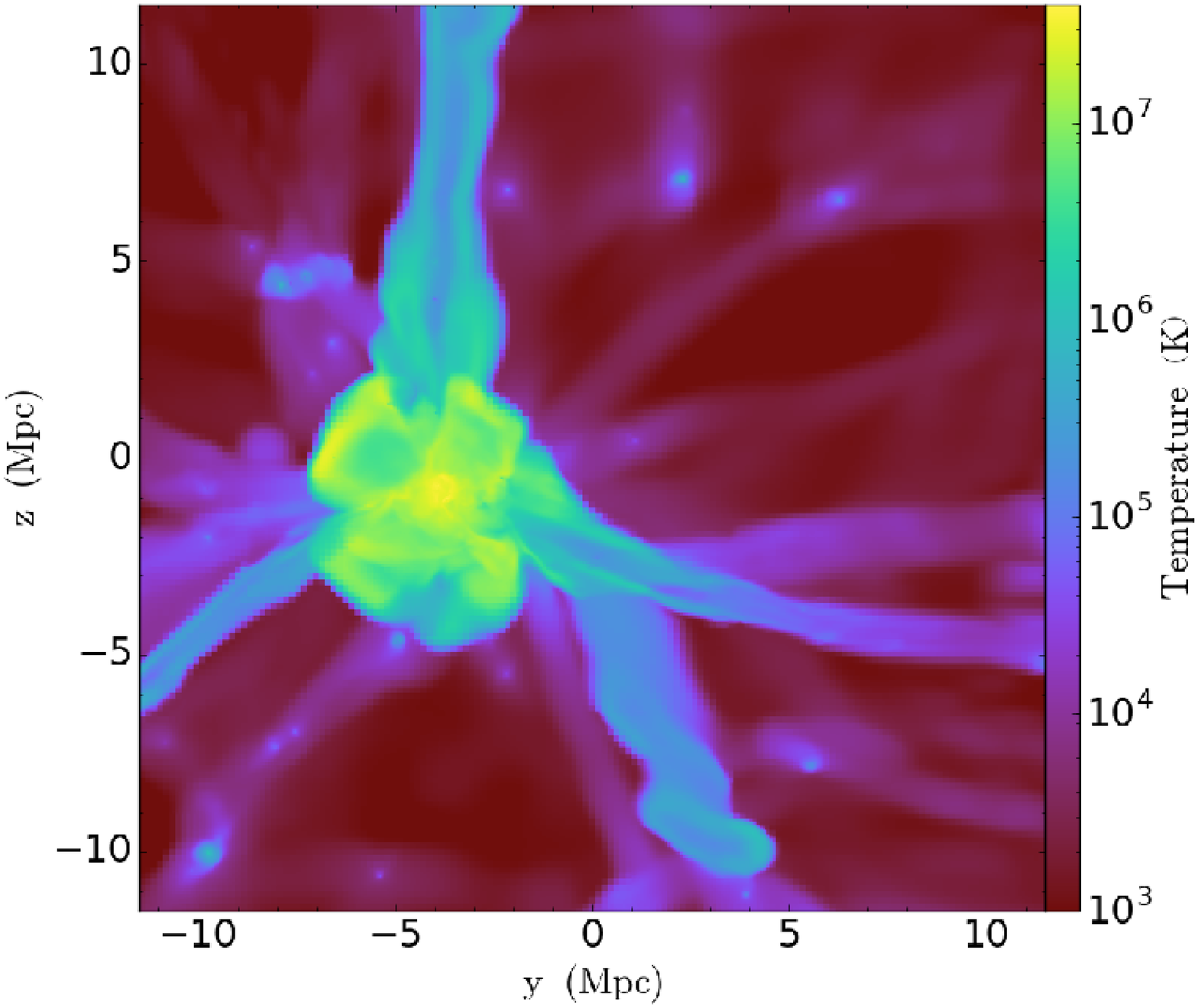}%{CThinxp5_p1wsy_wd_sf_Projection_x_temperature_dens.eps}
\includegraphics[scale=0.37, trim=40mm 5mm 5mm 0mm,clip]{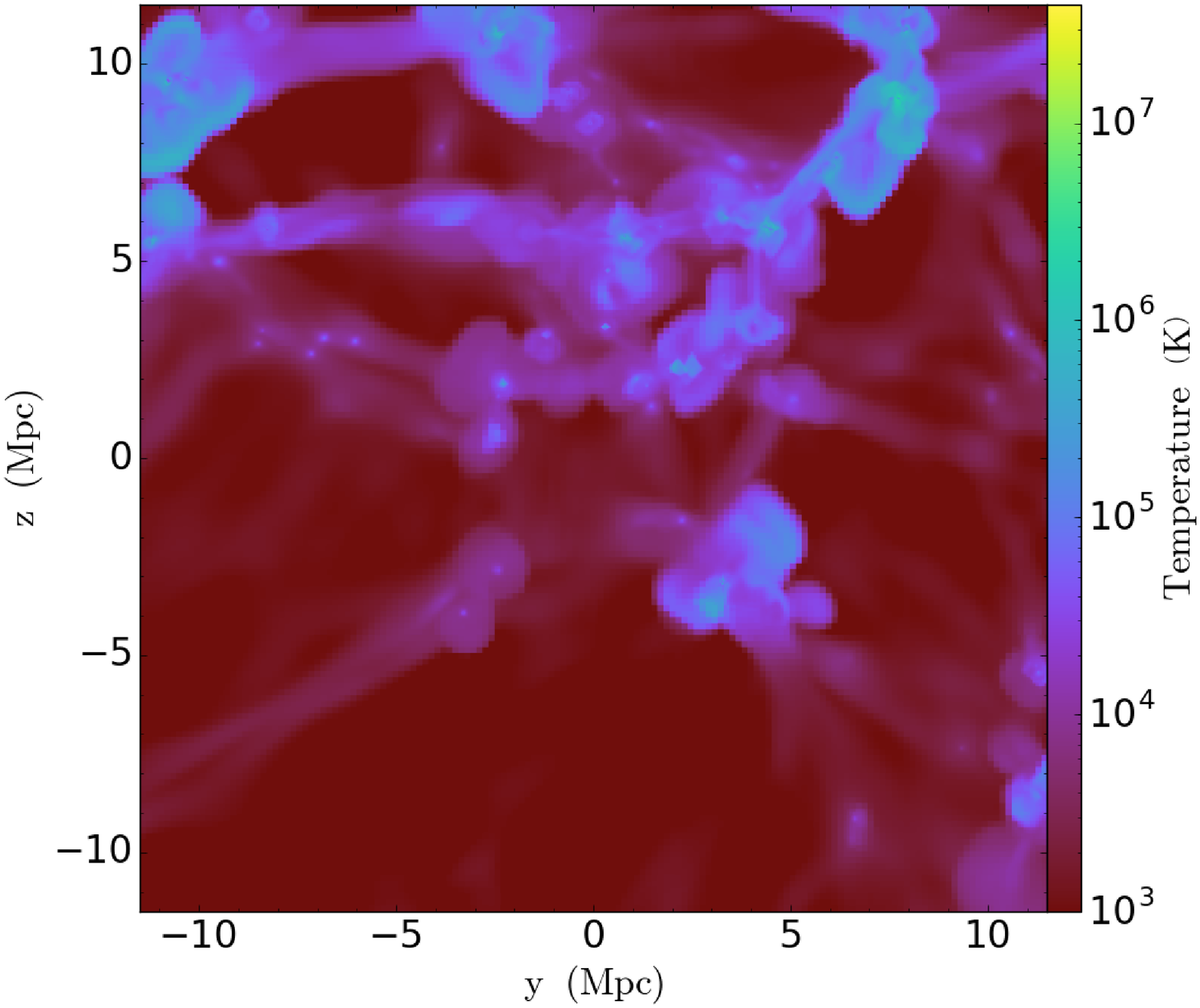}\\%{VThinxp5_p1wsy_wd_sf_Projection_x_temperature_dens.eps}\\
\includegraphics[scale=0.37,trim=10mm 5mm 49mm 0mm, clip]{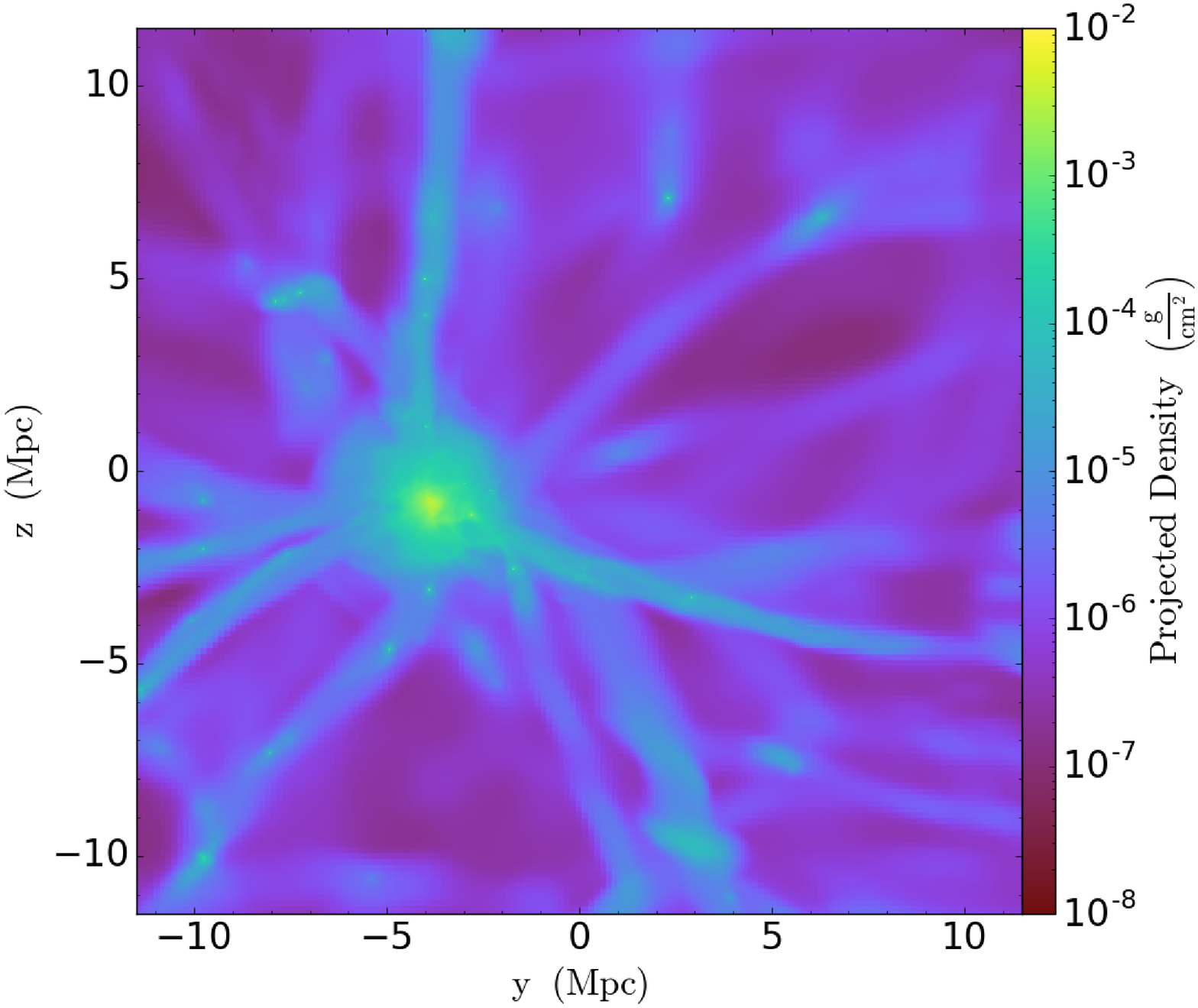}%{CThinxp5_p1wsy_sf_Projection_x_density.eps}
\includegraphics[scale=0.37, trim=40mm 5mm 5mm 0mm,clip]{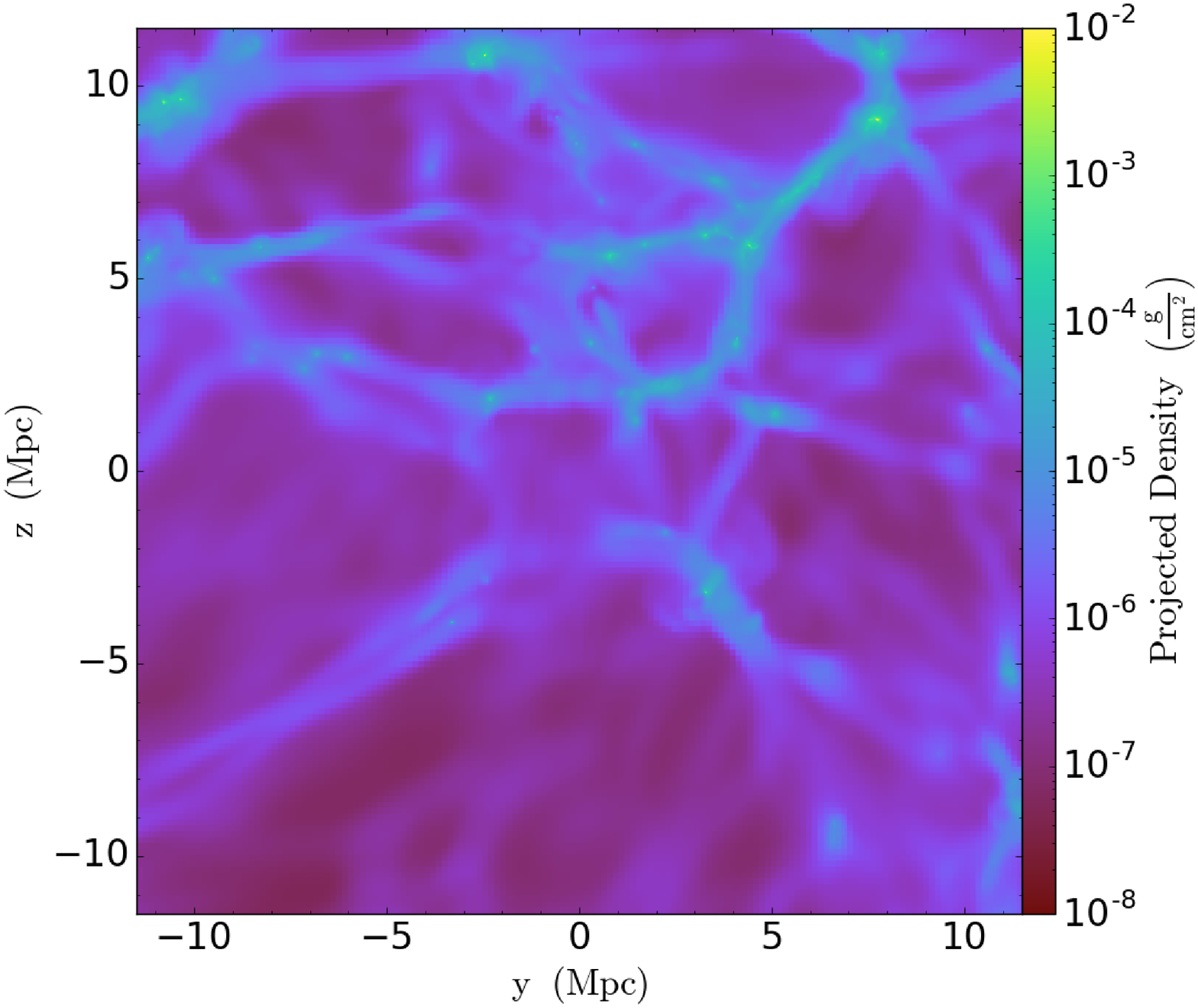}\\%{VThinxp5_p1wsy_sf_Projection_x_density.eps}\\
\caption{ Projections of thin slabs of part of the refined regions.  The slabs are $\sim$1.5 Mpc deep and 23 Mpc on a side.  The right panels are of the overdense region and the left panels are of the underdense region.  \textbf{Top Panels:}  Projected temperature weighted by density.  \textbf{Second Panels:}  Projected density of all gas.  \textbf{Third Panels:}  Projected HI fraction of the gas weighted by density.  \textbf{Bottom Panels:}  Projected HI number density.  The number densities shown span the column density of absorbing clouds we discuss in this paper. \label{fig:slices}}
\end{figure*}
\renewcommand{\thefigure}{\arabic{figure} (Cont.)}
\addtocounter{figure}{-1}
\begin{figure*}[!htb]
\includegraphics[scale=0.37,trim=10mm 5mm 49mm 0mm, clip]{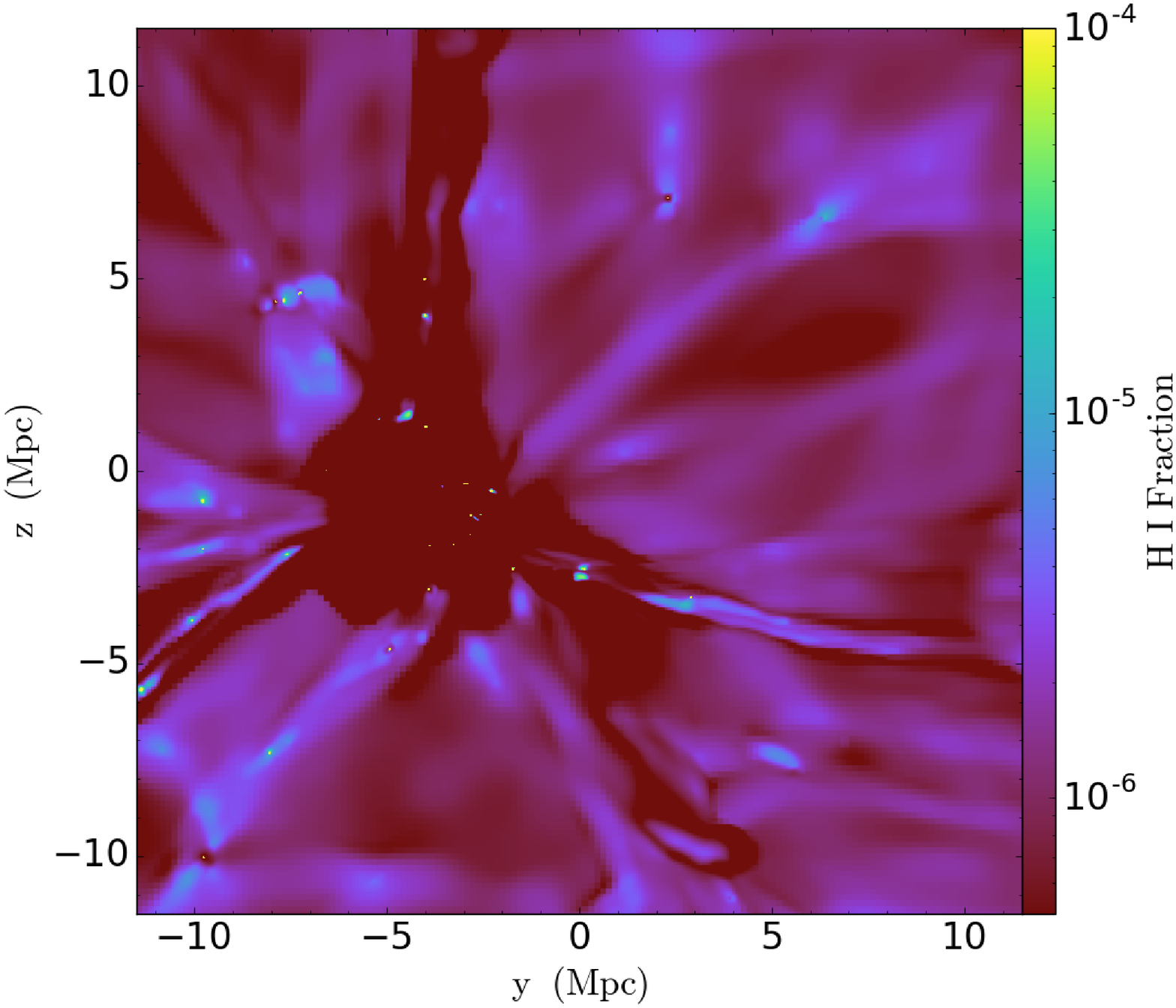}%{CThinxp5_p1wsy_wd_sf_Projection_x_H_p0_fraction_de.eps}
\includegraphics[scale=0.37, trim=40mm 5mm 5mm 0mm,clip]{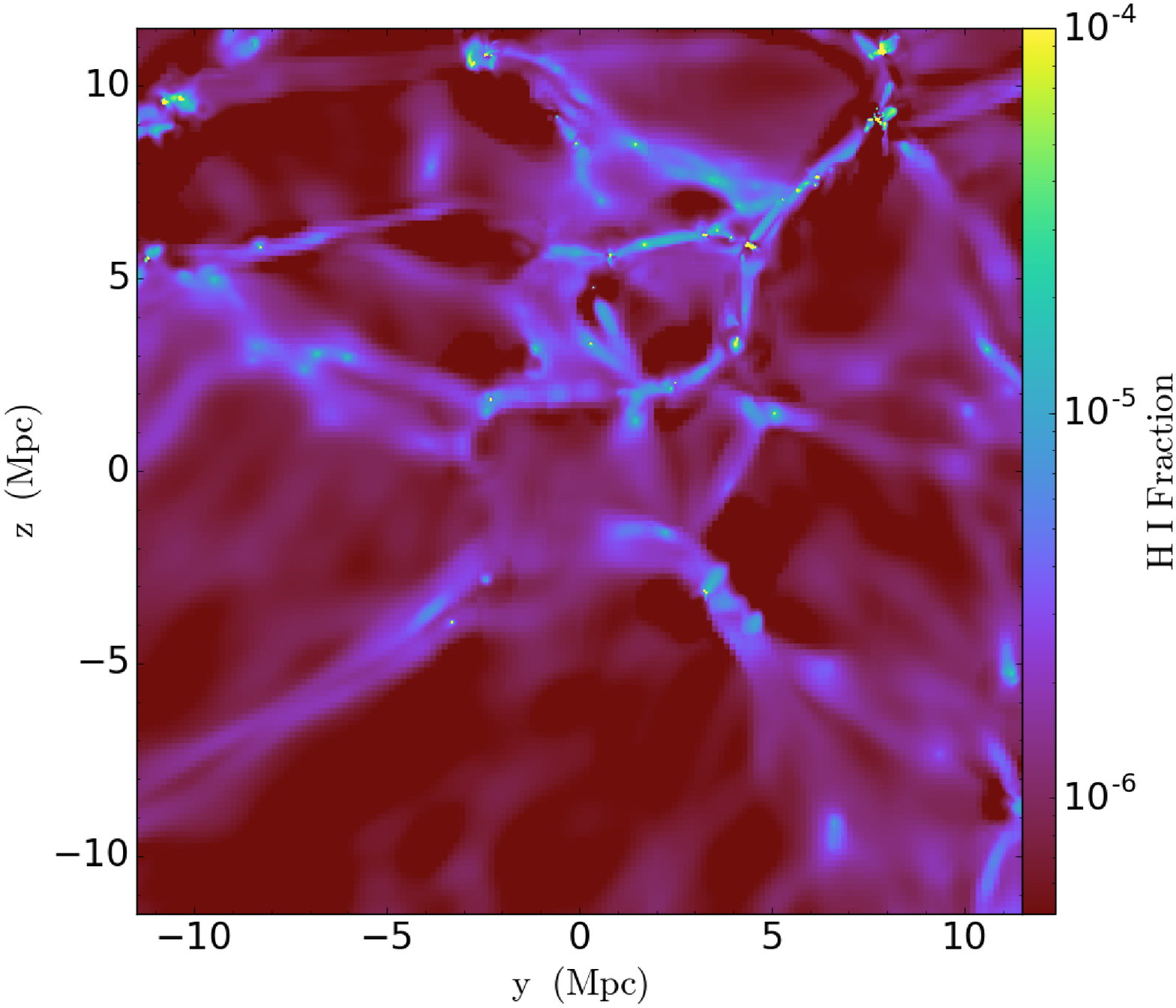}\\%{VThinxp5_p1wsy_wd_sf_Projection_x_H_p0_fraction_de.eps}\\
\includegraphics[scale=0.37,trim=10mm 5mm 49mm 0mm, clip]{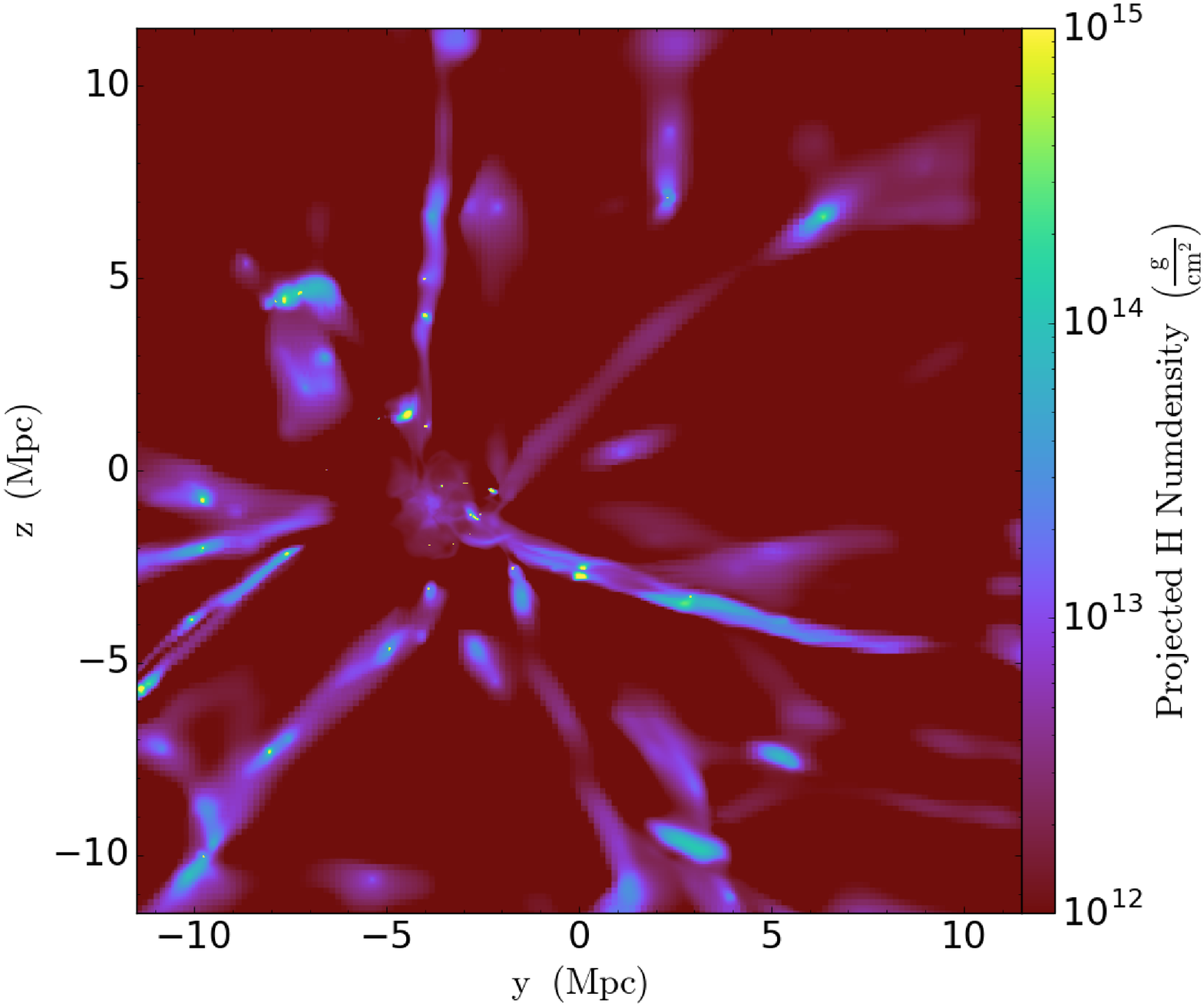}%{CThinxp5_p1wsy_sf_Projection_x_H_numdensity.eps}
\includegraphics[scale=0.37, trim=40mm 5mm 5mm 0mm,clip]{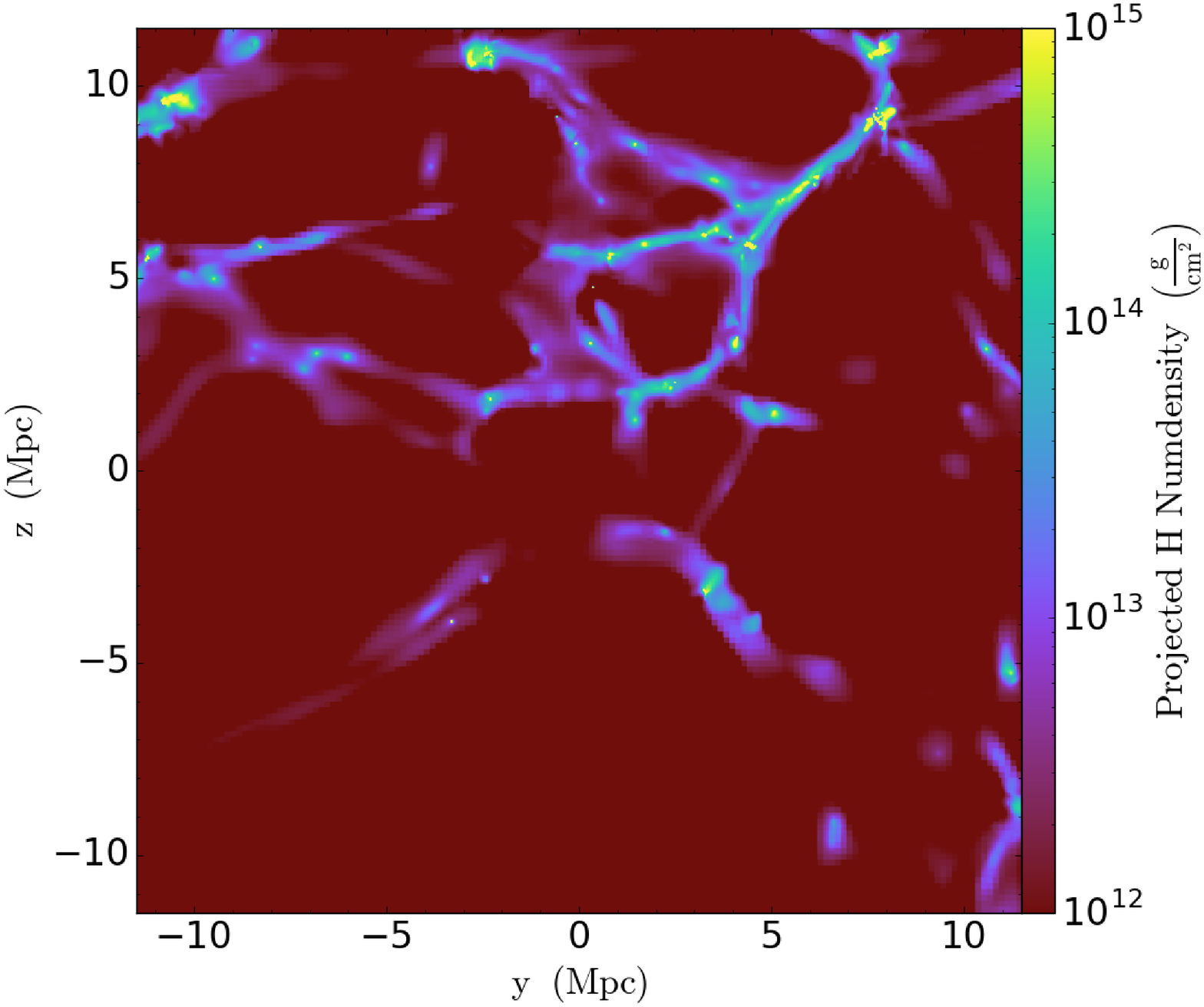}\\%{VThinxp5_p1wsy_sf_Projection_x_H_numdensity.eps}
\caption{Third and Fourth panels of Figure 8}
\end{figure*}
\renewcommand{\thefigure}{\arabic{figure}}

\section{Environmental Effects on Ly$\alpha$ Absorbers}\label{sec:cloudenvironment}

Why are absorbers in the overdense and underdense environments different?  We have found that the most dramatic difference in absorbing clouds is that their temperature is higher in the overdense region than in the underdense region.  However, the low metallicity of absorbers and their large distance from nearby galaxies indicate that neither SN feedback nor halo accretion shocks account for this difference.  

In Figure \ref{fig:slices} we show slices from the overdense and underdense refined regions of our simulation in the left and right panels, respectively. The top panels are projections of gas temperature weighted by density, the second panels are of total gas column density, the third panels are of the HI fraction weighted by density, and the bottom panels are the HI column density.  The slices are about 1.5 Mpc in width, which is somewhat larger than most of the Ly$\alpha$ absorbing clouds we find (Figure \ref{fig:Lmpc}).  

These slices are of a subsection of the refined regions, with 23 Mpc sides.  The slices highlight the cluster in the overdense region and contain the top of the void in the underdense region.  If we focus first on the cluster in the overdense region, we see that for this very massive halo the accretion shock has heated the gas to temperatures at which, despite a high gas column density, the HI column density is very low.  In general, we do not find Ly$\alpha$ absorbers around massive halos with a strong accretion shock, in agreement with our finding that absorbers tend to have temperatures below 10$^5$ K.  Instead, we find that high HI column density tends to resides in filaments (10$^{12.5}$ - 10$^{14.5}$ cm$^{-2}$).  In the temperature projection the filaments are quite wide, more than 1 Mpc, with temperature gradients that indicate shocking from structure formation.  The high density regions of the filaments are narrower, and only in the center, cooler regions is there significant HI column density.  

In the underdense region, we also see that much of the gas with HI column densities between 10$^{12.5}$ - 10$^{14.5}$ cm$^{-2}$ resides in filaments.  However, these filaments are much narrower in the temperature map, and rarely show signs of shocking in their temperature structure.  

We posit that the environment impacts Ly$\alpha$ absorbers in the formation of cool gas in filaments between galaxies.  Structure formation in the overdense region results in large collapsed structures outside of halos that have their own shocks that heat gas.  This hotter gas has lower HI fractions and therefore lower HI column density.

%fig9
\begin{figure*}
\includegraphics[scale=0.35,trim=10mm 5mm 49mm 0mm, clip]{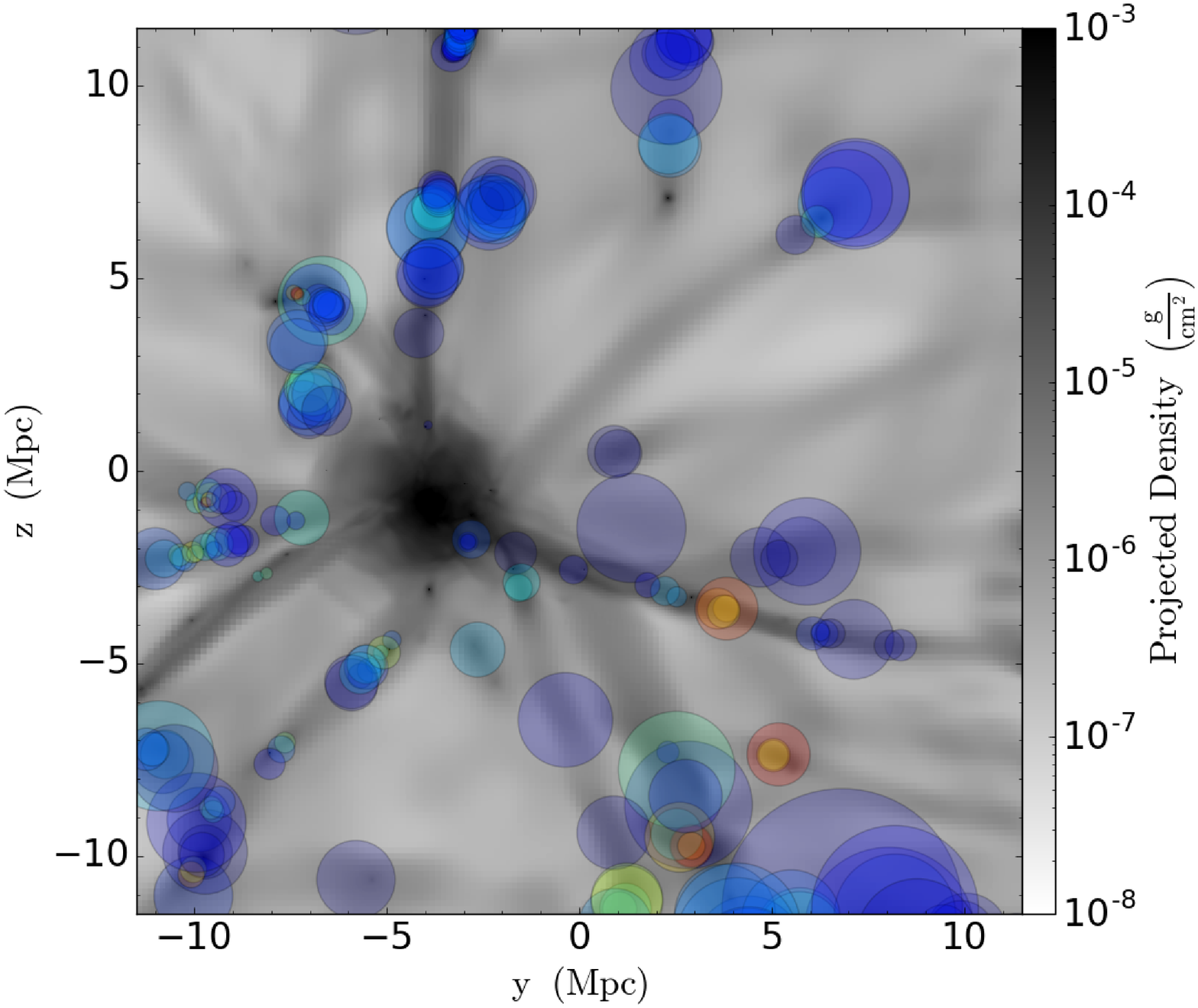}%{CThinxp5_p1wsy_sf_opc_patches_ncb.eps}%
\includegraphics[scale=0.35,trim=40mm 5mm 5mm 0mm, clip]{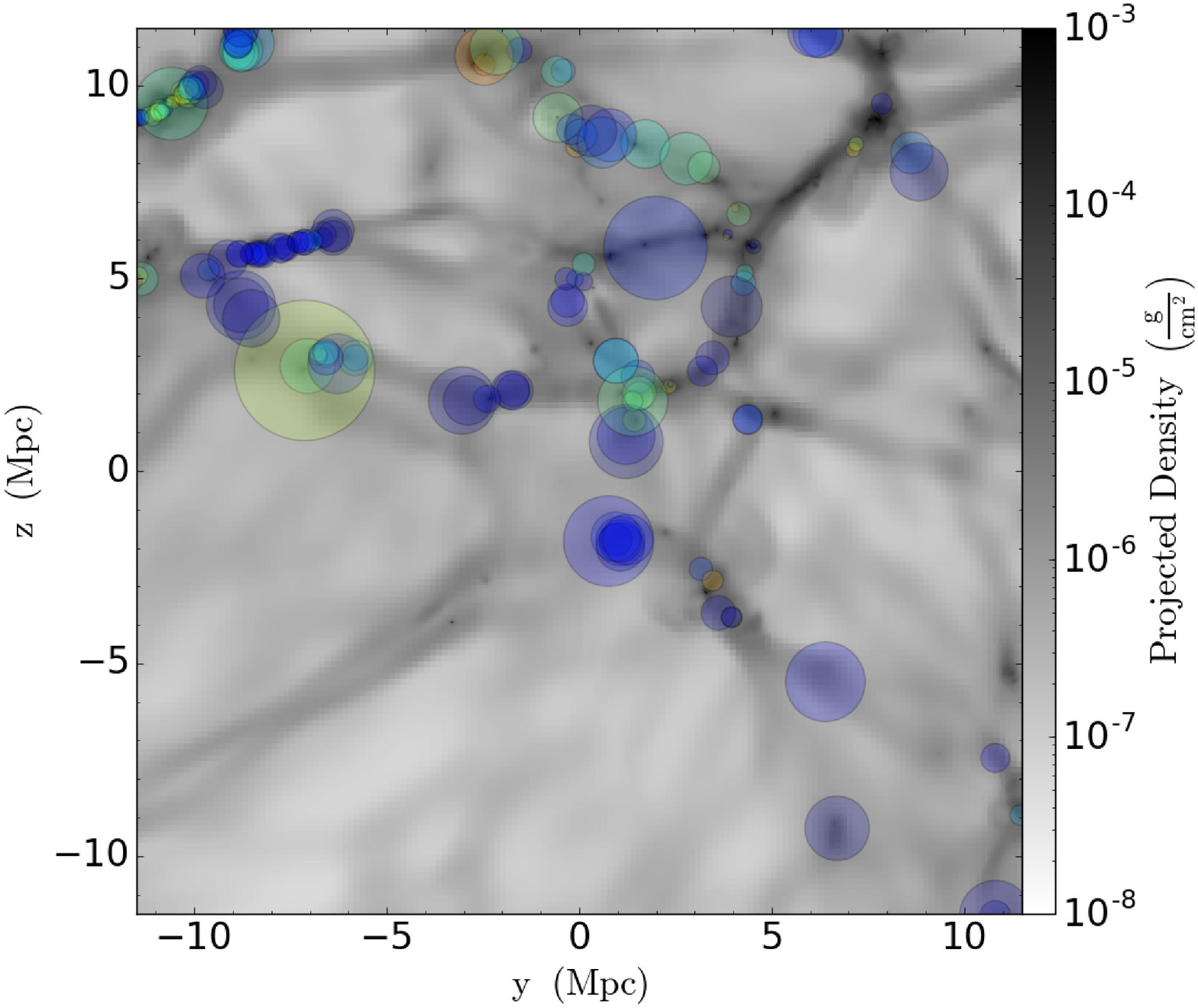}\\%{VThinxp5_p1wsy_sf_opc_patches_ncb.eps}\\%
\includegraphics[scale=0.38,trim=27mm 35mm 52mm 180mm, clip]{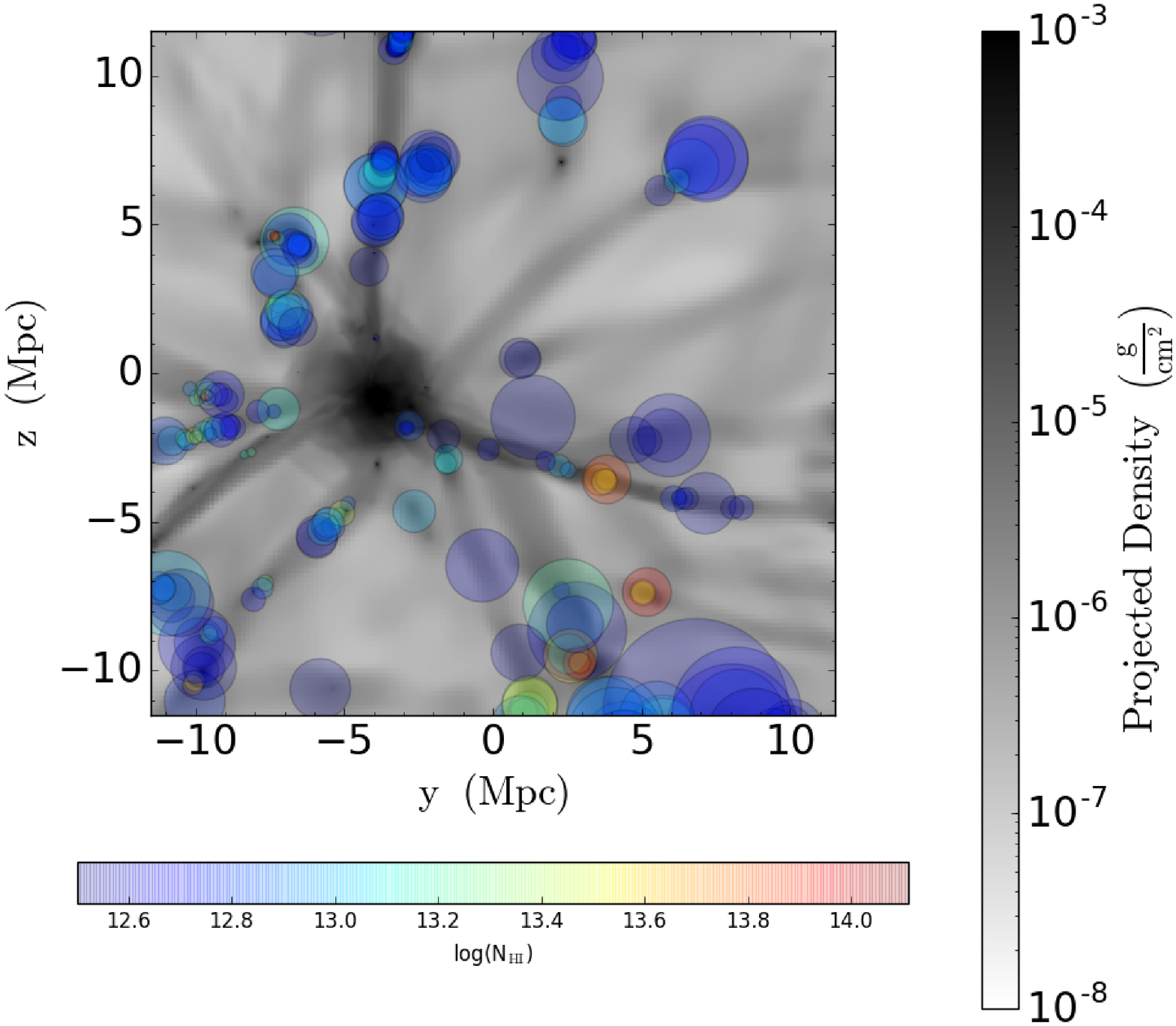}%{CThinxp5_p1wsy_sf_opc_patches.eps}
\includegraphics[scale=0.38,trim=40mm 35mm 52mm 180mm, clip]{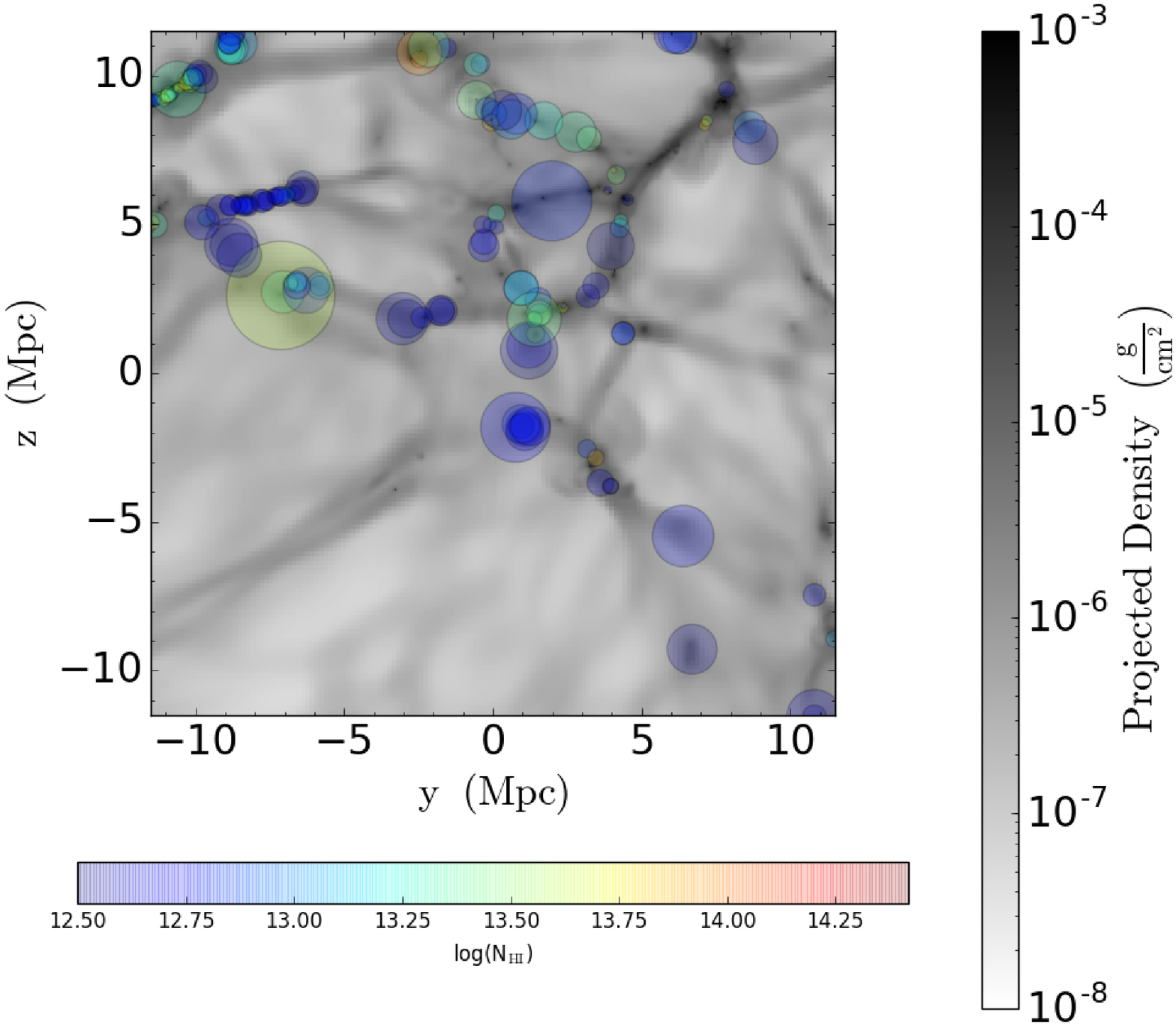}%{VThinxp5_p1wsy_sf_opc_patches.eps}

\caption{As in Figure \ref{fig:slices}, these are projections of the gas density in thin slabs of part of the refined regions.  The slabs are $\sim$1.5 Mpc deep and 23 Mpc on a side.  The right panel is of the overdense region and the left panel is of the underdense region.  The circles denote Ly$\alpha$ absorbing cloud positions whose peak HI density is within the central $\sim$300 pc of the slab.  The circle color and size denote the absorber column density and size, respectively.  \label{fig:sliceswithclouds}}
\end{figure*}

In Figure \ref{fig:sliceswithclouds} we illustrate this point using the narrow projection of gas density from Figure 8.  We have overplotted the positions of absorbing clouds whose HI peak density is within the central $\sim$300 pc of the slab (for clarity as the results do not change if we vary the line of sight range along which we select absorbers).  The colors of the spheres denote the column density of the absorbing cloud while the radius is half of the length of the absorber.  We see that several large, low-column density absorbing clouds lie along the comparatively hotter filaments in the overdense region, while the low-column density absorbers in the (cooler) filaments in the underdense region tend to be smaller.

This difference in absorbers is above and beyond any differences that would be driven by UV background sources.  In fact, UV sources might vary with environment in a way that is likely to exacerbate the differences between overdense and underdense environments.  For example, quasars tend to reside in higher density regions, so would be more likely to ionize gas in the overdense region.  Because the overdense region has more galaxies, ionization from star formation and AGN feedback would be more likely to impact absorbers in overdense environments.  

\section{Conclusions}\label{sec:conclusions}

In this paper we have used a cosmological hydrodynamical simulation to examine the Ly$\alpha$ forest in detail.  We focused on two refined regions, an over- and underdensity, representing +1.8$\sigma$ and -1.0$\sigma$ fluctuations in order to determine if the environment affects the nature of the absorbing gas.  Our results are as follows:

1)  We find very good agreement between our spectral HI column density distribution (CDD) and those in K14, indicating that our results on the HI CDD 
are robust and that resolution and simulation code have little impact on the HI CDD (Section \ref{sec:HICDD}).

2)  The HI CDD in the overdense environment is steeper than in the underdense environment, which is most dramatically seen in the smaller number of high density absorbers in the overdense versus underdense environment (N$_{\rm HI}$ $\geq$ 10$^{13.7}$ cm$^{-2}$) (Figure \ref{fig:HI_CDD}).  

3)  We find that there are physical differences in individual Ly$\alpha$ absorbing clouds in the two environments, specifically that clouds in the overdense region are larger, hotter, and have lower HI fractions than those in the underdense region (Figures \ref{fig:Lmpc} - \ref{fig:temp}).

4)  In both environments, the metallicity of Ly$\alpha$ absorbing clouds is quite low, indicating that the gas was not recently expelled from galaxies.  At lower column densities (N$_{\rm HI}$ $<$ 10$^{14}$ cm$^{-2}$), the metallicity is lower in clouds in the overdense region than in clouds in the underdense region (Figure \ref{fig:metallicity}).

5)  Ly$\alpha$ absorbers tend to reside far from galaxies (Figure \ref{fig:gad}), and even high column density clouds (N$_{\rm HI}$ $\geq$ 10$^{14}$ cm$^{-2}$) are more than two virial radii from their nearest galaxy neighbor.  In fact, much of the Ly$\alpha$ forest resides along filaments between galaxies (Figure \ref{fig:slices}).

We conclude that the environmental difference in the HI CDD slopes is driven by the temperature differences of the absorbing clouds.  In the overdense region, the higher temperature and lower HI fraction means that clouds must be larger to have the same column density as clouds in the underdense region.  We find that Ly$\alpha$ absorbing clouds tend to be far from galaxies and reside in cool regions of filaments.  Because filaments are naturally narrow structures, absorbers cannot be arbitrarily large.  This results in a steeper HI CDD slope in the overdense region, which tends to require larger clouds.  

In earlier work, K14 termed the large difference between the the HI CDD predicted using HM12 on hydrodynamic cosmological simulations and the observed HI CDD the Photon Underproduction Crisis.  We reiterate that the mismatch between observations and simulations points to a substantial and exciting gap in our current understanding of the low-redshift universe.  As highlighted by Shull et a. (2015), a single UV background generically fails to reproduce a sufficiently large range of the HI CDD.  Multiple partial matches between theoretical and observational HI CDDs highlight the types of sources that must be observed in more detail and modeled with the highest fidelity possible.   

Both the amplitude and the slope of the HI CDD are important clues to the exchange of energy between galactic and intergalactic scales.  The HI CDD provides a key diagnostic of these largely degenerate models, and with further observational and theoretical constraints, can become even more useful.  For observers, constraining the SFRD and QSO luminosity functions to better than a factor of two will dramatically decrease the current leeway in simulations, as will a more well-defined escape fraction from galaxies as a function of redshift and galaxy mass.  Although we find that gravitational shocks are a stronger energy source heating the IGM than feedback at $z$$=$0 (also Cen \& Chisari 2011), simulators' continued efforts in correctly modeling the heating of gas through both stellar and AGN feedback, and the mixing of this gas into the IGM, is important for understanding IGM heating and ionization.  Radiative transfer codes will be important to determine the range of influence of QSOs.  We recommend that some simulations, where AGN feedback is implemented to be an important source of heat and ionizing photons even at $z$$=$0, be confronted with constraints provided by observations of Ly$\alpha$ absorbers in different environments, as we perform here, to understand their plausibility.  Finding agreement across a broad range of Ly$\alpha$ absorber column density (at least the entirety of 10$^{12.5}$ - 10$^{14.5}$ cm$^{-2}$) is another important assessment of how well simulations reproduce observations.  Knowing to what extent flux sinks as well as sources can effect the HI CDD is critically important for narrowing in on the precise magnitude and origin of heat and ionizing photons in the local universe. \\

\acknowledgments Computing resources were in part provided by the NASA High- End Computing (HEC) Program through the NASA Advanced Supercomputing (NAS) Division at Ames Research Center and in part by a grant from the Ahmanson Foundation.
The research is supported in part by NSF grants AST-1108700, AST15-15389 and NASA grant NNX12AF91G.  ST was supported by the Alvin E. Nashman Fellowship in Theoretical Astrophysics.\\

\end{document}